# Game Theory for Signal Processing in Networks

Giacomo Bacci, Samson Lasaulce, Walid Saad, and Luca Sanguinetti

### Abstract

In this tutorial, the basics of game theory are introduced along with an overview of its most recent and emerging applications in signal processing. One of the main features of this contribution is to gather in a single paper some fundamental game-theoretic notions and tools which, over the past few years, have become widely spread over a large number of papers. In particular, both strategic-form and coalition-form games are described in details while the key connections and differences between them are outlined. Moreover, a particular attention is also devoted to clarify the connections between strategic-form games and distributed optimization and learning algorithms. Beyond an introduction to the basic concepts and main solution approaches, several carefully designed examples are provided to allow a better understanding of how to apply the described tools.

## I. INTRODUCTION

game theory (GT) is a branch of mathematics that enables the modeling and analysis of the interactions between several decision-makers (called players) who can have conflicting or common objectives. A *game* is a situation in which the benefit or cost achieved by each player from an interactive situation depends, not only on its own decisions, but also on those taken by the other players. For example, the time a car driver needs to get back home generally depends not only on the route he/she chooses but also on the decisions taken by the other drivers. Therefore, in a game, the actions and objectives of the players are tightly *coupled*. Until very recently, GT has been used only marginally in signal processing (SP), with notable examples being some applications in robust detection and estimation [1] as well as watermarking [2] (in which the watermarking problem is seen as a game between the data embedder and the attacker). However, the real catalyzer of the application of GT to SP has been the blooming of all issues related to *networking* in general, and distributed networks, in particular. The interactions that take place in a network can often be modeled as a game, in which the network nodes are the players that compete or form coalitions to get some advantage and enhance their quality-of-service. The main motivation behind formulating a game in a network is the large interdependence between the actions of the network nodes due to factors such as the use of common resources (e.g., computational, storage, or spectral resources), with interference across wireless networks being an illustrative case study. Paradigmatic examples of this approach

G. Bacci was with the University of Pisa, Pisa, Italy, and is now with MBI srl, Pisa, Italy (gbacci@mbigroup.it). L. Sanguinetti is with the University of Pisa, Dipartimento di Ingegneria dell'Informazione, Pisa, Italy (luca.sanguinetti@unipi.it), and also with the Large Systems and Networks Group (LANEAS), CentraleSupélec, Gif-sur-Yvette, France (luca.sanguinetti@supelec.fr).

S. Lasaulce is with Laboratoire des Signaux et Systèmes, Supélec, Gif-sur-Yvette, France (lasaulce@lss.supelec.fr).

W. Saad is with Wireless@VT, Bradley Department of Electrical and Computer Engineering, Virginia Tech (walids@vt.edu). He is also an International Scholar, Dept. of Computer Engineering, Kyung Hee University, South Korea.

L. Sanguinetti is funded by the People Programme (Marie Curie Actions) FP7 PIEF-GA-2012-330731 "Dense4Green" and is also supported by the ERC Starting Grant 305123 MORE. W. Saad is supported, in part, by the U.S. National Science Foundation under Grants CNS- 1460316, CNS-1460333, CNS-1513697, and AST-1506297.

can be found in the broad field of SP for communication networks in which GT is used to address fundamental networking issues such as: controlling the power of radiated signals in wireless networks, with the line of research largely originated from the seminal work in [3]; beamforming for smart antennas [4]; precoding in multi-antenna radio transmission systems [5]; data security [6]; spectrum sensing in cognitive radio [7]; spectrum and interference management [8]; multimedia resource management [9]; and image segmentation [10], [11].

Spurred and motivated by the well-established application to the fields above, GT is also pervading many other branches of SP, and has very recently been used for modeling and analyzing the following "classical" SP problems: distributed estimation in sensor networks [12]; adaptive filtering [13]; waveform design for multiple-input multiple-output (MIMO) radar estimation [14]; jamming of wireless communications [15] and MIMO radar applications [16]; and finding the position of network nodes [17]. In addition to the examples above, we must eventually point out the important connection that is building up between GT and SP through the fields of machine learning algorithms [18] and distributed optimization [19]. As explained in Sec. III, there exists a tight relationship between game-theoretic concepts and *learning algorithm aspects*. In this respect, one of the key messages of this contribution is that the solution of a game (often called an equilibrium, as discussed later) can often coincide with the convergence point that results from the interaction among several automata that implement iterative or learning algorithms. Therefore, there is an important synergy between GT and the broad field of multi-agent learning.

Despite the clear intersection between GT, learning, and optimization, as corroborated by a significant number of SP papers which exploit GT, it is worth noting that games usually have some features that are not common in classical optimization problems. In this respect, GT possesses its own tools, approaches, and notions. For example, in contrast to a classical optimization problem in which a certain function must be optimized under some constraints, the very meaning of optimal decision, or, equivalently, strategy, is generally unclear in interactive situations involving several decision makers, since none of them controls all the variables of the problem and these players can also have different objectives. To address such situations, GT is enriched with concepts coming from different disciplines such as economics and biology. This leads to notions that one does not encounter when studying, for instance, convex optimization. Examples of these notions are auctions, cooperative plans, punishments, rationality, risk aversion, trembling hand, and unbeatable strategies, to name a few. Remarkably, such concepts can actually be exploited to design algorithms. Although a player can be an automaton, a machine, a program, a person, an animal, a living cell, a molecule, or more generally any decision-making entity, it is essential to have in mind that a game is first and foremost a mathematical tool, which aims at modeling and analyzing an interactive situation. Before delving into the specific details of the various game models, we first provide a detailed overview on the different game models available in the GT literature.

There are three dominant mathematical representations for a game: (*i*) the strategic form; (*ii*) the extensive form; and (*iii*) the coalition form. Other representations exist e.g., the standard form which is used in the theory of equilibrium selection [20], and the state-space representation [21] but their use is rather marginal. The extensive form, which is typically used to investigate dynamical situations in computer science, will not be discussed in this survey. The main reason is that the extensive form, although more general (see [22], [23] and references therein for more details) than the strategic form, is often mathematically less tractable for typical SP problems. Defining the corresponding model and providing important results related to the strategic form is the purpose of

Table I: List of acronyms

| | | | |
|---|---|---|---|
| BR | best response | OCF | overlapping coalition formation |
| BRD | best-response dynamics | PF | partition function |
| CCE | coarse correlated equilibrium | PO | Pareto optimality |
| CE | correlated equilibrium | PoA | price of anarchy |
| CF | characteristic function | RL | reinforcement learning |
| FP | fictitious play | RM | regret matching |
| NBS | Nash bargaining solution | SE | strong equilibrium |
| NE | Nash equilibrium | SO | social optimality |
| NTU | non-transferable utility | TU | transferable utility |

Sec. II, whereas Sec. III shows how some solution concepts that are inherent to the strategic form can be related to algorithmic aspects. Sec. IV discusses the *coalition form*, which, unlike the strategic form, deals with options available to subsets of players (called cooperative groups or coalitions), what cooperative coalitions can form, and how the coalition utility is divided among its members. The algorithms that can be used to implement this approach are detailed in Sec. V. Note that, as described throughout the paper, for a given SP problem, the structure of the problem at hand and the practical constraints associated with it will determine whether the strategic or the coalition form is the most suitable representation. For example, it may occur that both forms are acceptable in terms of information assumptions, while complexity issues will lead to selecting one over the other.

To sum up, the main objectives of this tutorial are as follows. The primary goal of this survey is to provide a holistic reference on the use of GT in SP application domains. Some surveys have already been published in the SP literature [24] and communications literature [25] and [26]. The authors' motivation is not only to provide a refined and updated view of GT with respect to these existing tutorials, but also to establish explicit connections across the different tools of GT. This tutorial is intended for researchers and graduate students (with some expertise in networks and SP) interested in obtaining a comprehensive overview of game-theoretic concepts and distributed algorithm design,[1] and aims to:

- give the reader a global – although necessarily partial – overview of GT highlighting connections and differences between strategic-form and coalition-form games in a single paper;
- delineate differences and connections between GT and optimization;
- explain the strong relationship between game-theoretic solution concepts, such as the Nash equilibrium, and distributed SP algorithms;
- provide many application examples to help the reader understanding the way the described tools can be applied to different contexts.

For the reader's convenience, Fig. 1 provides a reference for the structure of this tutorial, adopting the typical methodology used to address game-theoretic problems and listing the topics described in each section, whereas

---

[1]For absolute beginners in GT, we refer readers to a recent lecture note [27], whereas we invite those interested in a thorough and textbook-oriented discussion on GT applied to wireless communications and SP to refer to the specific textbooks [22], [23].

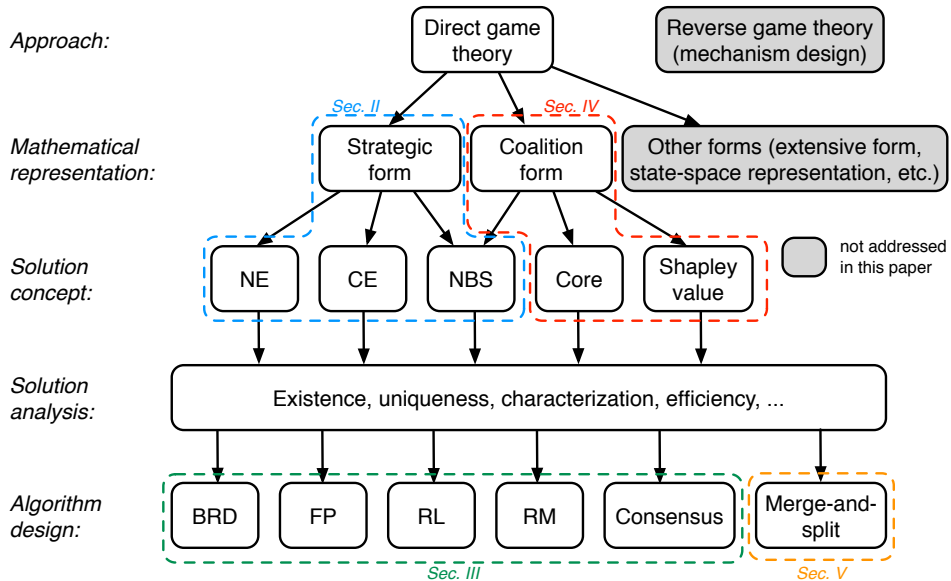

Fig. 1: Logical structure of the tutorial.

Table I lists the acronyms for the game-theoretic terms used throughout the tutorial.

## II. STRATEGIC-FORM GAMES

### A. Definition

A game in strategic (equivalently, normal) form is represented by a family of multi-variate functions $u_1, \ldots, u_K$; $K \geq 1$. The index set of this family, which is denoted here by $\mathcal{K} = \{1, \ldots, K\}$, is called the *set of players* and, for each $k \in \mathcal{K}$, $u_k$ is commonly called the *utility* (equivalently, *payoff*) *function* of player $k$. The strategic form assumes that $u_k$ can be any function of the following form:

$$u_k: \quad \begin{aligned} \mathcal{S}_1 \times \ldots \times \mathcal{S}_K &\rightarrow \mathbb{R} \\ (s_1, \ldots, s_K) &\mapsto u_k(s) \end{aligned} \tag{1}$$

where $\mathcal{S}_k$ is called the *set of strategies* of player $k$, $s_k$ is the strategy of player $k$, $s = (s_1, \ldots, s_K) \in \mathcal{S}$ is the strategy profile, and $\mathcal{S} = \mathcal{S}_1 \times \ldots \times \mathcal{S}_K$. We refer to a strategic-form game by using the compact triplet notation $\mathcal{G} = (\mathcal{K}, (\mathcal{S}_k)_{k \in \mathcal{K}}, (u_k)_{k \in \mathcal{K}})$. The notation $s_{-k} = (s_1, \ldots, s_{k-1}, s_{k+1}, \ldots, s_K)$ is used to denote the strategies taken by all other players except player $k$. With a slight abuse of notation, the whole strategy profile is denoted by $s = (s_k, s_{-k})$. The strategic-form representation may encompass a large number of situations in SP. To mention a few examples, players in a game can be: radars competing to improve their performance in terms of probability of false alarm or miss detection; sensors in a sensor network, which coordinate to estimate a field in a distributed way; base stations allocating the resources in a cellular network to optimize the system throughput; several digital signal processors, which have to compete for or manage computing resources; a watermarking device or algorithm, which has to find a good strategy against potential attackers.

Formally, it is worth noting that, in its general formulation, the strategic form is characterized by the simultaneous presence of two key features:

- each player $k$ can have its own objective, which is captured by a per-player specific function $u_k(s)$;

- each player $k$ has partial control over the optimization variables as it can control its strategy $s_k \in \mathcal{S}_k$ only.

Although the first feature is tied with *multi-objective optimization*, a clear difference exists in the control of the optimization variables, as in multi-objective optimization one has full control over all the variables.[2] Additionally, quite often in multi-objective optimization problems (see for example [28]), an aggregate objective must be defined. The second feature is tightly related to the framework of *distributed optimization*, although a common objective function is usually considered in this context, i.e., $\forall k \, u_k(s) = u(s)$. More importantly, the conventional assumption in distributed optimization is that the decision-making process is basically driven by a single designer (controller), which provides a set of strategies that the players strictly follow. Although being a possible scenario (which might be very relevant for some algorithmic aspects), in GT the players in general can also have the freedom to choose their strategies by themselves.

A central question is how to "solve" a strategic-form game. The very notion of optimality in this context is unclear since, as explained previously, we are in the presence of multiple objectives and the variables, which impact the utility functions, cannot be controlled jointly. This is the reason why the problem needs to be defined before being solved and why there exists the need for introducing game-theoretic *solution concepts*.

### B. Solution concepts

The Nash equilibrium (NE) is a fundamental solution concept for a strategic-form game on top of which many other concepts are built. This section is mostly dedicated to the NE and discusses more briefly other solution concepts, which might also be considered. In [29], Nash proposed a simple but powerful solution concept, which is now known as an *NE* (equivalently, *Nash point*).

**Definition 1** (Nash equilibrium)**.** *An NE of the game* $\mathcal{G} = (\mathcal{K}, (\mathcal{S}_k)_{k \in \mathcal{K}}, (u_k)_{k \in \mathcal{K}})$ *is a strategy profile* $s^{\mathrm{NE}} = (s_1^{\mathrm{NE}}, \ldots, s_K^{\mathrm{NE}}) = (s_k^{\mathrm{NE}}, s_{-k}^{\mathrm{NE}})$ *such that:*

$$\forall k \in \mathcal{K}, \forall s_k \in \mathcal{S}_k, \ \ u_k \left( s_k^{\mathrm{NE}}, s_{-k}^{\mathrm{NE}} \right) \geq u_k \left( s_k, s_{-k}^{\mathrm{NE}} \right). \tag{2}$$

A simple instance of an NE in everyday life would be to say that if everyone drives on the right, no single driver has an interest in driving on the left. As a more technical comment on the above definition, it can be seen that $s^{\mathrm{NE}}$ represents a strategy profile in the broad sense. For instance, it may be a vector of actions, a vector of probability distributions, or a vector of functions. Probability distributions naturally appear when considering an important extended version of the strategies of $\mathcal{G}$, namely *mixed strategies*. When $\mathcal{S}_k$ is finite,[3] they are defined as follows.

**Definition 2** (Mixed strategies)**.** *Let* $\Delta(\mathcal{X})$ *be the set of distribution probabilities over the generic set* $\mathcal{X}$ *(that is the unit simplex). Player $k$'s mixed strategy* $\pi_k \in \Delta(\mathcal{S}_k)$ *is a distribution that assigns a probability* $\pi_k(s_k)$ *to each strategy* $s_k$, *such that* $\sum_{s_k \in \mathcal{S}_k} \pi_k(s_k) = 1$. *For mixed strategies, the (joint) probability distribution over the strategy profile $s$ is by definition the product of the marginals* $\pi_k$, $k \in \mathcal{K}$.

---

[2]In fact, one can also define a strategic-form game in which a player has multiple objectives.

[3]The continuous case is obtained by using an integral instead of a discrete sum in the definition.

A mixed strategy thus consists in choosing a lottery over the available actions. In the case where a player has two possible choices, choosing a mixed strategy amounts to choosing a coin with a given probability of having head (or tail): the player flips the coin to determine the action to be played. Using mixed strategies, each player can play a certain strategy $s_k$ with probability $\pi_k(s_k)$. Note that the strategies considered so far, termed *pure strategies*, are simply a particular case of mixed strategies, in which probability 1 is assigned to one strategy, and 0 to the others. The importance of mixed strategies, aside from being more general mathematically than pure strategies, comes in part from the availability of existence results for mixed NE. The latter is defined as follows.

**Definition 3** (Mixed Nash equilibrium)**.** *A mixed strategy NE of the game* $\mathcal{G} = (\mathcal{K}, (\mathcal{S}_k)_{k \in \mathcal{K}}, (u_k)_{k \in \mathcal{K}})$ *is a mixed strategy profile* $\pi^{\mathrm{NE}} = (\pi_1^{\mathrm{NE}}, \ldots, \pi_K^{\mathrm{NE}}) = (\pi_k^{\mathrm{NE}}, \pi_{-k}^{\mathrm{NE}})$ *such that*

$$\forall k \in \mathcal{K}, \forall \pi_k \in \Delta(\mathcal{S}_k), \widetilde{u}_k\left(\pi_k^{\mathrm{NE}}, \pi_{-k}^{\mathrm{NE}}\right) \geq \widetilde{u}_k\left(\pi_k, \pi_{-k}^{\mathrm{NE}}\right) \tag{3}$$

*where*

$$\widetilde{u}_k\left(\pi_k, \pi_{-k}\right) = \mathbb{E}(u_k) = \sum_{s \in \mathcal{S}} \left(\prod_{j \in \mathcal{K}} \pi_j(s_j)\right) u_k\left(s\right) \tag{4}$$

*is the expected utility of player* $k$ *when selecting the mixed strategy* $\pi_k$*, and* $\mathcal{S} = \mathcal{S}_1 \times \ldots \times \mathcal{S}_K$*.*

By definition, an NE of $\mathcal{G}$ is a point such that, for every index $k$, the function $u_k$ cannot be (strictly) increased by just changing the value of the variable $s_k$ at the equilibrium. For this reason, an NE is said to be strategically stable to unilateral deviations. The NE has at least two other very attractive features:

- In its mixed version, its existence is guaranteed for a broad class of games;
- It may result from the repeated interaction among players, which are only partially informed about the problem. In particular, some well-known distributed and/or learning algorithms may converge to an NE (see Section III).

Elaborating more on the first feature, it should be stressed that existence is a fundamental issue in GT. In fact, one might think of various solution concepts for a game. For example, one might consider a point which is stable to $K$ deviations rather than to a single one (with $K$ being the number of players). This solution concept is known as a *strong equilibrium (SE)* (e.g., see [22], [23]): an SE is a strategy profile from which no group of players (of any size) can deviate and improve the utility of every member of the group while the players outside the deviating group maintain their strategy to that of the equilibrium point. The SE is therefore stable to multiple deviations and the number of them can be up to $K$. This is a strong requirement, which explains why it is quite rarely satisfied in a static game (see [22] for a static-game example where it is met). In fact, the SE is particularly relevant in infinitely repeated games. To better understand this, the reader is referred to Sec. IV where the notion of core is described; indeed, it turns out that a specific version of the core, named the $\beta$-core, of a game coincides with the SE utilities in an infinite repetition of that game [30]. Considering the SE as a solution concept (in a context of purely selfish players of a static game) might be inappropriate since it will typically not exist, instead, the NE offers more positive results in terms of existence. Indeed, tackling the existence issue of an NE for a strategic-form game $\mathcal{G}$ reduces to study a fixed-point problem for which quite positive results can be obtained. To this end, the notion of *best-response (BR)* for a player must be first introduced.

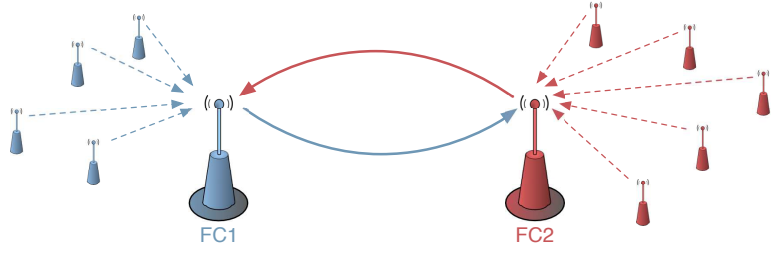

Fig. 2: The wireless sensor's dilemma.

**Definition 4** (best-response). *Player $k$'s best-response $\mathrm{BR}_k(s_{-k})$ to the vector of strategies $s_{-k}$ is the set-valued function*

$$\mathrm{BR}_k(s_{-k}) = \arg \max_{s_k \in \mathcal{S}_k} u_k(s_k, s_{-k}).$$ (5)

By introducing the auxiliary notion of composite (or, equivalently, global, game's) best-response

$$\begin{aligned} \mathrm{BR}: \quad \mathcal{S} \quad &\rightarrow \quad\quad\quad\quad \mathcal{S} \\ s \quad &\mapsto \quad \mathrm{BR}_1(s_{-1}) \times \ldots \times \mathrm{BR}_K(s_{-K}), \end{aligned}$$ (6)

we have the following characterization for an NE.

**Definition 5** (NE characterization). *Let $\mathcal{G} = (\mathcal{K}, (\mathcal{S}_i)_{i \in \mathcal{K}}, (u_i)_{i \in \mathcal{K}})$ be a strategic-form game. A strategy profile $s^{\mathrm{NE}}$ is an NE if and only if:*

$$s^{\mathrm{NE}} \in \mathrm{BR}(s^{\mathrm{NE}}).$$ (7)

The characterization of an NE in terms of a fixed-point problem is due to Nash [29] and explains why common existence theorems are based on topological and geometrical assumptions such as compactness for the sets of strategies or continuity for the utility functions. The following two theorems explain why the NE is an attractive solution concept from the *existence* issue standpoint: they show that any finite game or compact continuous game possesses at least one mixed NE.

**Theorem 1** ([31]). *In a strategic-form game $\mathcal{G} = (\mathcal{K}, (\mathcal{S}_k)_{k \in \mathcal{K}}, (u_k)_{k \in \mathcal{K}})$, if $\mathcal{K}$ is finite and $\mathcal{S}_k$ is finite for every $k$, then there exists at least one NE, possibly involving mixed strategies.*

**Theorem 2** ([31]). *In a strategic-form game $\mathcal{G} = (\mathcal{K}, (\mathcal{S}_k)_{k \in \mathcal{K}}, (u_k)_{k \in \mathcal{K}})$, if $\mathcal{S}_k$ is compact and $u_k$ is continuous in $s \in \mathcal{S}$ for every $k \in \mathcal{K}$, then there exists at least one NE, possibly involving mixed strategies.*

To better picture out the meaning of the strategic-form representation and the notion of NE, let us consider a simple example, which is an instance of what is referred to as the prisoner's dilemma in the GT literature [32].

**Example 1** (The wireless sensor's dilemma). *Consider the wireless sensor network sketched in Fig. 2, which is populated by a number of wireless sensors sending their own measurements (e.g., target detection, temperature), to their fusion centers (FCs), labeled as* FC1 *and* FC2. *For the sake of graphical representation, sensors communicating with sensors, and the FCs themselves, are represented with blue and red colors, respectively.*

Fig. 3: A wireless sensor's dilemma game under matrix form.

As is known, gathering information at each FC from a larger population of nodes (in this case, those covered by the other FC) helps improving its measurement accuracy. However, sharing data among different population of nodes implies additional transmission of information across the FCs, which is in general costly due to energy expenditure. In this context, the two FCs can independently and simultaneously decide whether to share (i.e., relay) the information or not. Depending on both decisions, each FC gets a (dimensionless) utility in the form "accuracy minus spent energy", given according to Fig. 3 (known as payoff matrix), in which $0 \leq e \leq 1$ represents the cost incurred by an FC for relaying the measurements of the other.

The communication problem corresponding to Example 1 can be modeled as a strategic-form game where the set of players is $\mathcal{K} = \{\text{FC1, FC2}\}$ and the action (strategy) sets are $\mathcal{S}_k = \{\text{sleep mode, active mode}\}$ for $k \in \{1, 2\}$. The utility function for FC1 (the one for FC2 follows by symmetry) is given by:

$$u_1(s_1, s_2) = \begin{vmatrix} -e & \text{if} & (s_1, s_2) & = & (\text{active mode, sleep mode}) \\ 0 & \text{if} & (s_1, s_2) & = & (\text{sleep mode, sleep mode}) \\ 1-e & \text{if} & (s_1, s_2) & = & (\text{active mode, active mode}) \\ 1 & \text{if} & (s_1, s_2) & = & (\text{sleep mode, active mode}). \end{vmatrix} \tag{8}$$

This game is said to be a *static* (equivalently, *one-shot*) *game* since each player takes a single action once and for all. Since this game is finite, it has at least one mixed NE (according to Theorem 1). To find these equilibria, let us denote by $\rho_1$ (resp., $\rho_2$) the probability that FC1 (resp., FC2) assigns to the action active mode. The mixed NE of the considered game can be found by computing the expected utilities. For player FC$k$ with $k \in \{1, 2\}$, it writes as $\widetilde{u}_k(\rho_1, \rho_2) = -e\rho_k + \rho_{-k}$. The best-response of player $k$ is given by: $\forall \rho_{-k} \in [0, 1], \widetilde{\text{BR}}_k(\rho_{-k}) = 0$. Since, by definition, Nash equilibria are *intersection points* of the best-responses, the unique mixed NE is $(\rho_1^{\text{NE}}, \rho_2^{\text{NE}}) = (0, 0)$, which is a pure NE consisting of the action profile (sleep mode, sleep mode).

**Example 2** (The cognitive radio's dilemma)**.** *Observe that the example above is general enough to encompass many different applications. For example, it can be used to model a cognitive network with two cognitive radios (CRs), CR1 and CR2, which have to decide independently and simultaneously to transmit either over a narrow or a wide frequency band. In this case, the two corresponding actions are respectively denoted by* narrowband *and* wideband. *Depending on the CR's decisions, each CR transmits at a certain data rate (say in Mbps) accordingly to Fig. 4. The first (second) component of each pair corresponds to the transmission rate (i.e., utility) of* CR1

Fig. 4: A CR's dilemma game under matrix form (utilities may be expressed in Mbps).

(CR2). *For instance, if both use a wide band, their transmission rate is the same and equals* 1 *Mbps.*

The action (sleep mode) in Example 1 (or, in Example 2, wideband) is called a *strictly dominant action* for player $k$ since, for any given action chosen by the other player, it provides a strictly higher utility than any other choice. At the equilibrium (sleep mode, sleep mode), the wireless sensors have a zero utility. We can see that there exists an action profile at which both players would gain a higher utility. The action profile (active mode, active mode) is said to *Pareto-dominate* the action profile (sleep mode, sleep mode). More generally, in any game, when there exists a strategy profile, which provides a utility for every player that is greater than the equilibrium utility, the equilibrium is said to be *Pareto-inefficient*. Inefficiency is generally a drawback of considering the NE as a solution concept. From an engineering point of view, it would be more desirable to find an equilibrium that is Pareto-efficient, i.e., a Pareto-optimal (PO) point.

**Definition 6** (Pareto-optimal profile). *A strategy profile $s^{\mathrm{PO}}$ is a PO point if there exists no other strategy profile $s$ such that $u_k(s) \geq u_k\left(s^{\mathrm{PO}}\right)$ for all $k \in \mathcal{K}$, and $u_k(s) > u_k\left(s^{\mathrm{PO}}\right)$ for some $k \in \mathcal{K}$.*

In addition to Pareto optimality, another related concept widely used is the weak PO point defined as follows:

**Definition 7** (weak PO profile). *A strategy profile $s^{\mathrm{PO}}$ is a weak PO point if there exists no other strategy profile $s$ such that $u_k(s) > u_k\left(s^{\mathrm{PO}}\right)$ for all $k \in \mathcal{K}$.*

In other words, when operating at a PO strategy profile, it is not possible to increase the utility of one player without decreasing that of at least one other. In many occasions, beyond the concept of Pareto optimality, the performance (in terms of social efficiency) of an NE can be measured by comparing it to a socially optimal profile, which is defined as a maximizer of the *social welfare*[4] (or, more properly, sum-utility) $\sum_{k \in \mathcal{K}} u_k(s)$. Formally stated, a social-optimal (SO) point is defined as follows.

**Definition 8** (social optimum). *A strategy profile $s^{\mathrm{SO}}$ is a social optimum point if it satisfies*

$$s^{\mathrm{SO}} \in \arg\max_{s \in \mathcal{S}} \sum_{k \in \mathcal{K}} u_k(s). \tag{9}$$

[4]Other global measures can be used to introduce some fairness (e.g., see [33]). Through Definition 12, the Nash product (defined later on in this section) is considered and can be shown to be proportionally fair (see [34]).

Fig. 5: A simple CR's coordination game which exhibits non-trivial CE (utilities may be expressed in Mbps).

Both PO and SO points can be seen as possible solution concepts for a game. Often, implementing these solution concepts will require some coordination between the players, and typically rely on the need for significant information and knowledge assumptions. In the framework of distributed networks, such coordination degree and/or knowledge might not be available or may be costly, and, thus, social and Pareto optimality can only be used to measure the performance loss induced by decentralization. There is a common and simple measure of efficiency, which allows us to quantify the gap between the performance of centralized (in some sense, classical) optimization and distributed optimization. Indeed, the efficiency of the Nash equilibria can be measured using the concept of *price of anarchy (PoA)* [35], which is defined as follows.

**Definition 9** (price of anarchy). *The PoA corresponds to*

$$\text{PoA} = \frac{\max\limits_{s \in \mathcal{S}} \sum\limits_{k \in \mathcal{K}} u_k(s)}{\min\limits_{s \in \mathcal{S}^{\text{NE}}} \sum\limits_{k \in \mathcal{K}} u_k(s)} \tag{10}$$

*where $\mathcal{S}^{\text{NE}}$ denotes the set of all NE in a game.*

Otherwise stated, the PoA provides a measure of the performance loss (in terms of social welfare) of the "worst" NE compared to a socially optimal strategy. The closer the PoA to 1, the higher the efficiency of the NE. One of the features of PoA is that it can be upper bounded in some important cases, e.g. in congestion games with monomial costs [36]; a *congestion game* is a special form of games in which the utility of a player depends on its own action and depends on others' action only through the way they distribute over the available actions (often called edges or routes). For instance, if the cost (the opposite of the utility) is linear, the PoA is upper bounded by $\frac{4}{3}$, showing that the price of decentralization is relatively small in this scenario.

To illustrate the notions of PoA, let us reconsider Example 2, where the four possible utility profiles are reported in Fig. 4. The game has three Pareto optima: $(1, -e)$, $(-e, 1)$, and $(1 - e, 1 - e)$. Geometrically, a utility vector is PO if there is no point in the North-East orthant whose origin is located at the candidate point. In the considered game, there is a unique NE. Here, the PoA equals $\frac{3+3}{1+1} = 3$. If there is no means of coordinating the two CRs, which may happen when both transmitters have been designed independently or are owned by different economic players, the loss in terms of social efficiency has to be undergone. However, if there is a common designer as in the framework of distributed optimization, it may be possible to decrease the PoA.

**Remark 1.** *One way to improve efficiency is to keep on considering an NE as the solution concept but to transform the game. The corresponding general framework is referred to as* mechanism design *[37]. Affine pricing is a very special instance of mechanism design: it consists in applying an affine transformation on the utility functions and tune the introduced parameters to obtain an NE, which is more efficient than the one considered in the original game [3].*

Another possibility to improve efficiency is to keep the game unchanged but to modify the solution concept. This may be either a *CE* or a *Nash bargaining solution (NBS)*. A CE is a joint distribution over the possible actions or pure strategy profiles of the game from which no player has interest in deviating unilaterally. More formally, we have the following definition.

**Definition 10** (correlated equilibrium)**.** *A CE is a joint probability distribution* $q^{\mathrm{CE}} \in \triangle(\mathcal{S})$ *which verifies:*

$$\forall k \in \mathcal{K}, \forall s'_k \in \mathcal{S}_k, \sum_{s_{-k} \in \mathcal{S}_{-k}} q^{\mathrm{CE}}(s_k, s_{-k}) u_k(s_k, s_{-k}) \geq \sum_{s_{-k} \in \mathcal{S}_{-k}} q^{\mathrm{CE}}(s'_k, s_{-k}) u_k(s'_k, s_{-k}), \tag{11}$$

*where* $\mathcal{S}_{-k} = \mathcal{S}_1 \times \ldots \times \mathcal{S}_{k-1} \times \mathcal{S}_{k+1} \times \ldots \times \mathcal{S}_K$.

We know that a pure NE is a special case of mixed NE for which the individual probability distributions used by the players are on the vertices of the unit simplex. We see now that a mixed NE is a special case of a CE for which joint probability distributions over the action profiles factorizes as the product of its marginals. One important question is to know how to obtain a CE in practice. Aumann showed that the availability of a "exogenous public signal" to players allows the game to reach new equilibria, which are in the convex hull of the set of mixed NE of the game [22]. By "public signal", it is implied that every player can observe it; the adjective "exogenous" is added to explicitly indicate that the signal is not related to the player's actions. A simple example would be the realization of a Bernouilli random variable such as the outcome obtained by flipping a coin. Additionally, if exogenous private signals are allowed, new equilibria outside this hull can be reached and lead to better outcomes; by "private" it is meant that each player observes the realizations of its own lottery. The obtained equilibria are precisely CE. Having a CE therefore means that the players have no interest in ignoring (public or private) signals, which would recommend them to play according to the realizations of a random lottery whose joint distribution corresponds to a CE $q^{\mathrm{CE}}$. In the case of the wireless sensor's dilemma, it can be checked that the only CE boils down to the unique pure NE of the game, showing that sending a broadcast signal to the wireless sensors would not allow them to reach another equilibrium, which might be more efficient. To better picture out the meaning of CE, consider the modified version of Example 2 shown in Fig. 5 under matrix form. Observe that it has not the structure of a prisoner's dilemma anymore (no strictly dominant strategy for the players exists). Fig. 6 shows the set of CE of this game. In particular, it turns that a public signal allows the CR to reach any CE in the convex hull of the points $(5, 1)$, $(1, 5)$.

Another notion of equilibrium derived from the notion of CE is the *coarse correlated equilibrium (CCE)*. It is mathematically more general than the CE, and hence the set of CE is included in the set of CCE. One of the motivations for mentioning it here is that CCE can be learned by implementing simple algorithms such as regret matching based learning schemes [18] (see Sec. III for further details).

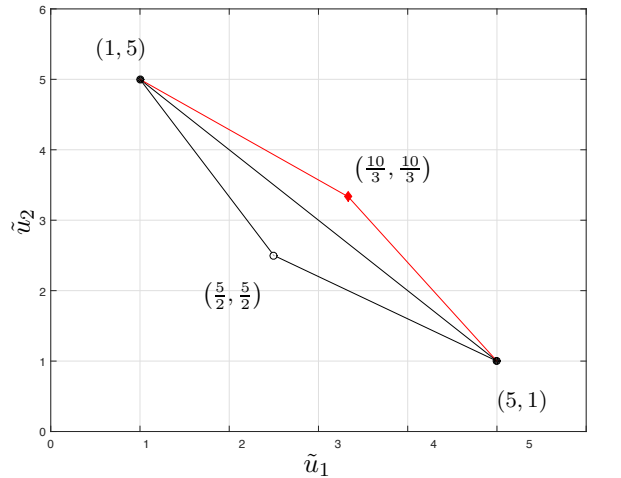

Fig. 6: Set of CE of the game given by Fig. 5 in the expected utility plane. A public signal allows the CR to reach any CE in the convex hull of the points $(5,1)$, $(1,5)$ (pure NE) and $\left(\frac{5}{2}, \frac{5}{2}\right)$ (strict mixed NE). Private signals allow one to extend this region. The set of CE becomes the convex hull of the points $(1,5)$, $(5,1)$, $\left(\frac{5}{2}, \frac{5}{2}\right)$, and $\left(\frac{10}{3}, \frac{10}{3}\right)$.

**Definition 11** (coarse correlated equilibrium). *A CCE is a joint probability distribution* $q^{\mathrm{CCE}} \in \triangle(\mathcal{S})$ *which verifies:*

$$\forall k \in \mathcal{K}, \forall s'_k \in \mathcal{S}_k, \ \sum_{s \in \mathcal{S}} q^{\mathrm{CCE}}(s) u_k(s) \geq \sum_{s_{-k} \in \mathcal{S}_{-k}} q^{\mathrm{CCE}}_{-k}(s_{-k}) u_k(s'_k, s_{-k}) \tag{12}$$

*where* $q^{\mathrm{CCE}}_{-k}(s_{-k}) = \sum_{s'_k \in \mathcal{S}_k} q^{\mathrm{CCE}}(s'_k, s_{-k})$.

A possible interpretation of this definition is as follows. Following the notion of CCE, players are assumed to decide, before receiving the recommendation associated with a public or private signal, whether to commit to it or not. At a CCE, all players are willing to follow the recommendation given that all the others also choose to commit. That is, if a single player decides not to follow the recommendations, it experiences a lower (expected) utility. Based on this interpretation, the difference between the CCE and the CE is that in the latter, players choose whether or not to follow a given recommendation *after* it has been received. Therefore, there is no a priori commitment.

Another effective approach that can be taken to further improve the efficiency of the game solution while also addressing fairness issues is to seek alternative solution concepts. One example of such alternative solutions is the concept of NBS [38], which has been originally defined for two-player games. The implementation of the NBS typically requires some form of coordination or exchange of information among the players. As explained in Sec. III, the NBS can be related to SP algorithms such as *consensus algorithms*. The NBS has been used in the networking literature about twenty years ago to obtain fair solutions to flow control problems in communication networks [33]. More recently, it has been exploited in different contexts such as: in [34] to solve bandwidth allocation problems; in [39] to achieve weighted proportional fairness in resource allocation in wireless networks; or, in [4] to obtain cooperative beamforming strategies in interference networks where transmitters are equipped with multiple antennas. Another example can be found in [9] wherein the bargaining methodology is employed to address the problem of rate allocation for collaborative video users (see also [40]). Following [38], let us define the NBS for two-player games. For this, we denote by $\mathcal{U}$ the set of feasible utility points of the strategic-form game of interest and assume that $\mathcal{U}$ is a closed and convex set. Let denote by $(\lambda_1, \lambda_2)$ a given point in $\mathcal{U}$, which will be referred to as a status quo (or, equivalently, disagreement point). The NBS is then defined as follows.

**Definition 12** (Nash bargaining solution). *The NBS is the unique PO profile, which is a solution of*

$$\max_{(u_1, u_2) \in \mathcal{U}} \quad (u_1 - \lambda_1)(u_2 - \lambda_2) \tag{13}$$

$$\text{subject to} \quad u_1 \geq \lambda_1, \ u_2 \geq \lambda_2.$$

The graphical interpretation of the NBS is shown in Fig. 7. The solution of (13) corresponds to the point of tangency between the Pareto boundary of $\mathcal{U}$ and the hyperbola $(u_1 - \lambda_1)(u_2 - \lambda_2) = \kappa$, where $\kappa$ is properly chosen to ensure only one intersection between the two curves. The original definition of the NBS by Nash only concerns two-player games but it can be extended by considering $K$ players. For this, the two-factor product above, which is called the *Nash product*, becomes $\prod_{k=1}^{K} (u_k - \lambda_k)$. However, when there are more than two parties involved in the bargaining, coalition forming is always possible and this definition may need to be replaced by modified versions, such as the coalition NBS [41] (please refer to Sec. IV for further details). We will conclude the discussion on the NBS by providing an example which is drawn from [42], namely a beamforming game for communications in presence of interference.

**Example 3** (Beamforming game [42]). *Consider two $N$-antenna transmitters. Transmitter $i \in \{1, 2\}$ has to choose a beamforming vector $w_i \in \mathbb{C}^N$ such that $w_i^H w_i = 1$ (where the superscript $^H$ stands for Hermitian transpose). The signal observed by the single-antenna receiver $i$ is given by $y_i = h_{ii}^H w_i x_i + h_{ji}^H w_j x_j + z_i$, $j = -i$, $h_{ji} \in \mathbb{C}^N$ are fixed for all $(i, j)$, $x_i \in \mathbb{C}$, and $z_i \sim \mathcal{CN}(0, 1)$ is a complex white Gaussian noise. By choosing the utility function as $u_i = \log(1 + \text{SINR}_i)$ with $\text{SINR}_i = \frac{|h_{ii}^H w_i|^2 \mathbb{E}|x_i|^2}{1 + |h_{ji}^H w_j|^2 \mathbb{E}|x_j|^2}$, it can be shown that any point of the Pareto frontier can be reached by beamforming vectors which linearly combine the zero-forcing (ZF) beamforming solution $(w_i^{\text{ZF}})$ and maximum ratio transmission beamforming (MRT) solution $(w_i^{\text{MRT}})$ [42]. Therefore, finding the NBS amounts to finding the appropriate linear combination coefficient $\alpha_i$ which is defined as $w_i = \alpha_i w_i^{\text{ZF}} + (1 - \alpha_i) w_i^{\text{MRT}}$. The unique NE of the considered game corresponds to $(\alpha_1^{\text{NE}}, \alpha_2^{\text{NE}}) = (0, 0)$ that is each transmitter uses ZF beamforming. By choosing the unique NE of the game under investigation to be the status quo point, i.e., $\lambda_i = u_i(\alpha_i^{\text{NE}}, \alpha_j^{\text{NE}})$, the NBS is then given by:*

$$(\alpha_1^{\text{NBS}}, \alpha_2^{\text{NBS}}) = \arg \max_{(\alpha_1, \alpha_2) \in [0, 1]^2} [u_1(\alpha_1, \alpha_2) - u_1(0, 0)] \times [u_2(\alpha_1, \alpha_2) - u_2(0, 0)]. \tag{14}$$

*By construction, the obtained solution is necessarily more Pareto-efficient than the NE. However, computing the NBS typically requires more channel state information than what is required by the NE [42].*

### C. Special classes of strategic-form games

In this subsection, we review some special classes of strategic-form games, that show a relevant share of the game-theoretic approaches available in the SP literature. For the sake of brevity, we list here only the distinguishing features of each class but provide also a (non-exhaustive) list of relevant references that can be used to gather more specific details on problem modeling and solution tools. For other interesting classes of games (not reported here due to space constraints), the interested readers are referred to specific literature on the topic (e.g., [22], [23], [31]).

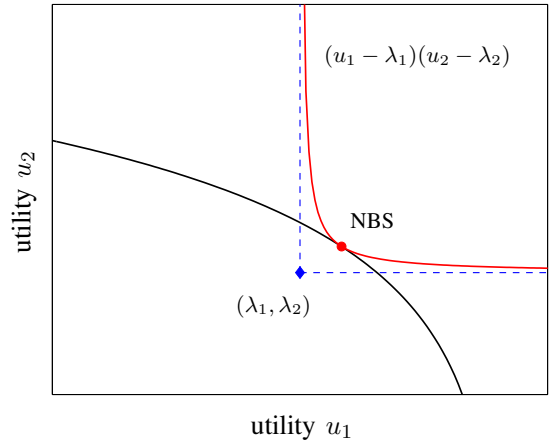

Fig. 7: Graphical interpretation of the NBS point (red circle) as the intersection between the Pareto boundary of $\mathcal{U}$ and the hyperbola $(u_1 - \lambda_1)(u_2 - \lambda_2) = \kappa$, where the status quo $\lambda = (\lambda_1, \lambda_2)$ is represented by the blue diamond.

*Zero-sum games:* One of the most common types of strategic-form games is the two-player zero-sum game. A two-player zero-sum game is a game in which the sum of the utilities is zero or can be made zero by appropriate positive scaling and translation which do not depend on the player's actions or strategies. In other words, it is a game such that $\mathcal{K} = \{1, 2\}$, $u_1(s_1, s_2) + u_2(s_1, s_2) = 0$. In such a game, one player is a maximizer, i.e., aims to maximize its gain, while the other player is a minimizer, i.e., aims to minimize its losses (which are the gains of the other player). In SP, zero-sum games are especially popular when modeling security games involving an attacker and a defender. In such games, the attacker's gains are most often equal to the defender's losses, yielding a zero-sum situation. An example in this context can be found in [16] in which the interaction between a target and a MIMO radar – both smart – is modeled as a two-player zero-sum game since the target and the radar are completely hostile. The mutual information criterion is used in formulating the utility functions. In [43], the problem of polarimetric waveform design for distributed MIMO radar from a game-theoretic perspective is also formulated as a two player zero-sum game played between an opponent and the radar design engineer. In [2], the authors use a two-player zero-sum game to model a watermarking problem where a source sequence (the *covertext*) needs to be copyright-protected before it is distributed to the public. Another example is given by a two-user communication channel (such as the Gaussian multiple access channel) with a constraint on the total sum-rate [44].

Despite being one of the most well studied and analyzed class of strategic-form games in GT (in part because many results can be derived), zero-sum games are restrictive. In fact, the majority of the studied problems in SP are better modeled as nonzero-sum games.

*Continuous quasi-concave games:* A game is said to be continuous if, for all $k \in \mathcal{K}$, the utility function $u_k$ is continuous in the strategy profile $s$. It is said to be quasi-concave if $u_k$ is quasi-concave w.r.t. $s_k$ for any fixed $s_{-k}$ and $\mathcal{S}_k$ is a compact and convex set. For such games, we can take advantage of Theorem 2, which ensures the *existence of at least one pure-strategy NE*. A flurry of research activity on energy-efficient resource allocation in wireless communications or sensor networks makes use of quasi-concave utility functions, that aim at trading off the performance of network agents while saving as much energy as possible. Since usually the performance is increasing with the amount of resources employed, a useful modeling provides

$$u_k(s) = \frac{f_k\left(\frac{s_k}{1+\sum_{j\neq k}s_j}\right)}{s_k} \tag{15}$$

under the hypothesis of a one-dimensional strategy set $\mathcal{S}_k = [0, P^{\max}]$, with $P^{\max}$ being the maximum transmit power. As long as $f_k$ shows some desirable properties (such as sigmoidness) which are often verified in many SP and communications scenario, the ratio $u_k$ proves to be quasi-concave w.r.t. $s_k$. This is the case for instance when $f(x) = (1 - e^{-x})^M$, $M \leq 1$ or $f(x) = e^{-\frac{a}{x}}$, $a > 0$.

*Continuous concave games:* The same assumptions as for the previous special class of games are made except that $u_k$ is now a concave function of $s_k$. The *existence of a pure NE* is guaranteed in such games since individual concavity implies individual quasi-concavity. Interestingly, if we make one more assumption, called the diagonally strict condition (DSC), *uniqueness* of the NE can also be guaranteed. This is worth mentioning since sufficient conditions for ensuring uniqueness are quite rare in the literature of GT. The DSC is met if there exists a vector of (strictly) positive components $r = (r_1, \ldots, r_K)$ such that:

$$\forall (s, s') \in \mathcal{S}^2, s \neq s': \ (s - s')\left(\gamma_r(s') - \gamma_r(s)\right)^T > 0 \tag{16}$$

where $\gamma_r(s) = \left[r_1 \frac{\partial u_1}{\partial s_1}(s), \ldots, r_K \frac{\partial u_K}{\partial s_K}(s)\right]$. An example of this game can be found in [45]. Therein, the scenario investigated is a set of multi-antenna transmitters which have to choose a precoding matrix to optimize their expected individual transmission rate between each of them and a common multi-antenna receiver.

*Super-modular games:* Super-modular games are thoroughly investigated in [46]. A strategic-form game is super-modular if, for all $k \in \mathcal{K}$, $\mathcal{S}_k$ is a compact subset of $\mathbb{R}$; $u_k$ is upper semi-continuous in $s$; and $u_k(s_k, s_{-k}) - u_k(s_k, s'_{-k})$ is non-decreasing in $s_k$ for all $k \in \mathcal{K}$, $\mathcal{S}_k$ and for all $s_{-k} \geq s'_{-k}$, where the inequality is intended to be component-wise. In the example of power control, this definition is very easy to understand. If all the transmitters, except $k$, increase their power level, then transmitter $k$ has interest in increasing its own power as well. Two properties make super-modular games appealing in the SP community: *i)* the set of *pure-strategy NE* is *not empty*; and *ii)* iterative distributed algorithms such as the best-response dynamics (BRD) (see Sec. III for more details) can be used to let the players *converge* to the one of the NE of the game. As an example, perform an affine transformation of the utility functions in (15) such that they become

$$U_k(s) = \frac{f_k\left(\frac{s_k}{1+\sum_{j\neq k}s_j}\right)}{s_k} - c_k s_k \tag{17}$$

with $c_k \geq 0$ being a parameter to be tuned. The latter parameter induces a penalty in terms of utility which increases with the transmit power. The corresponding transformation is called affine or linear pricing and aims at improving (social) efficiency at the equilibrium. The corresponding game can be shown to be super-modular provided that the action space is reduced as detailed in [3]. Other examples of super-modular games can be found in the literature of SP. For instance, in [17] the problem of time of arrival-based positioning is formulated as a super-modular game.

*Potential games:* A strategic-form game is said to be potential if, for all $k \in \mathcal{K}$, $s_k, s'_k \in \mathcal{S}_k$ and all $s_{-k} \in \mathcal{S} \setminus \mathcal{S}_k$, the difference $u_k(s_k, s_{-k}) - u_k(s'_k, s_{-k})$ can be related to a global *potential function* $\Phi(s)$ that does not depend on the specific player $k$. There exist at least four types of potential games: weighted, exact, ordinal, and generalized ones, according to the relationship between the differences in utilities and potential functions [22]. For example, a game is an exact potential game if there exists a function $\Phi$ such that $u_k(s_k, s_{-k}) - u_k(s'_k, s_{-k}) = \Phi(s_k, s_{-k}) - \Phi(s'_k, s_{-k})$. Similarly to super-modular games, the interest in potential games stems from the guarantee of the *existence of pure-strategy NE*, and from the study of a single function, which allows the application of theoretical tools borrowed from other disciplines, such as convex optimization [47]. For instance, a maximum point for $\Phi$ is an NE for $\mathcal{G}$. Similarly to super-modular games, *convergence* of iterative distributed algorithms such as the BRD algorithm is guaranteed in potential games. Examples of potential games can be found in [48] for a problem of power allocation, in [49] for radar networks, or in [50] for a problem of multi-portfolio optimization. In [51], the authors make use of a potential game to study cooperative consensus problems for sensor deployment. Other simple examples of potential games are games with a common utility function or games for which each utility only depends on the individual action or strategy.

*Repeated games:* It is important to note that the definition of the strategic form does not require any particular assumption on the sets of strategies $\mathcal{S}_1, \ldots, \mathcal{S}_K$. In particular, as seen throughout this section, an example of $\mathcal{S}_k$ can be a discrete alphabet (as in the wireless sensor's dilemma), or an interval of $\mathbb{R}$ (as in the example of energy-efficient power control game). In the mentioned examples, the game is said to be static, because each player takes a single action. It should be stressed however that the strategic form can also be used to model some dynamic games in which players have to take an action in a repeated manner and even in a continuous-time manner (e.g., in some differential games). In dynamic games the sets of strategies become more complex objects. They can be sets of sequences of functions or sets of sequences of probability distributions. Due to space limitation, we will only mention the case of repeated games here, which will allow us to identify some differences in terms of modeling and analysis between static and repeated games.

A repeated game belongs to a subclass of dynamic games, in which the players face the same single-stage game, say $\Gamma = (\mathcal{K}, (\mathcal{A}_k)_{k \in \mathcal{K}}, (\nu_k)_{k \in \mathcal{K}})$, where $\mathcal{A}_k$ is the set of possible actions for player $k$, and $\nu_k$ is its instantaneous utility function. The game is played over several stages. The number of stages can be either finite or infinite. The single-stage game is called, equivalently, the *constituent*, *component*, or *stage game*. When introducing the notion of time, the strategies $s_k$ become complete plans of actions, that depend on the unfolding of the game through time. More precisely, a strategy in a repeated game typically corresponds to a sequence of maps or functions, which assign an action to a sequence of observations. Similarly, the utility functions of the repeated game are modified and correspond now to average or long-term utilities. Often, average utilities are of the form

$$u_k(s) = \sum_{t=1}^{+\infty} \theta_t \nu_k(a(t)), \tag{18}$$

where $(\theta_t)_{t \geq 1}$ represents a sequence of weights which can model different aspects depending on the scenario under consideration (e.g., see [22]). Typical choices for $(\theta_t)_{t \geq 1}$ are:

- $\forall t \in \{1, \ldots, T\}, \theta_t = \frac{1}{T}$ and $\forall t \geq T+1, \theta_t = 0$; this type of game is referred to as a finitely repeated game;
- $\forall t \geq 1, \theta_t = (1-\delta)\delta^t$ with $0 \leq \delta < 1$; this type of game is referred to as a repeated game with discount;
- when the limit exists, $\forall t \geq 1, \theta_t = \frac{1}{T}$; this type of game is called an infinitely repeated game.

The definition of the strategies $s_1, \ldots, s_K$ strongly depends on the observation assumptions made. For instance, in a repeated game with *perfect monitoring and perfect recall*, i.e., a game where every player observes all the past actions and is able to store them, the strategy of player $k \in \mathcal{K}$ is given by the following sequence of causal functions:

$$\forall t \geq 1, \ s_{k,t} \ : \quad \begin{aligned} \mathcal{A}^{t-1} &\rightarrow \mathcal{A}_k \\ (a(1), \ldots, a(t-1)) &\mapsto a_k(t) \end{aligned} \tag{19}$$

where $a(t) = (a_1(t), \ldots, a_K(t))$ is the profile of actions played at stage $t$ and $\mathcal{A}^0 = \emptyset$ by convention. This strategy is called a pure strategy.

Even in the special case of repeated games just described, we can identify some important differences between static and repeated games in terms of equilibrium analysis. The existence issue is fundamental for the NE to be relevant as a solution concept for the problem of interest. Note that, while uniqueness is an important issue for static games, e.g., to be able to predict the convergence point of a distributed algorithm, it is generally much less relevant for a repeated game, since the number of equilibria can be large and even infinite. This is the reason why equilibria are not characterized in terms of equilibrium strategies, but rather in terms of equilibrium utilities. This characterization corresponds to a theorem called the *Folk theorem* [31]. We have seen that efficiency is an important issue for a static game. For a repeated game, due to the fact that players can observe the history of the actions played and therefore exchange information, there may exist efficient equilibria and those equilibria can be attained. For example, in the case of the wireless sensor's dilemma, the following strategies can be checked to be equilibrium strategies of an infinite repeated game with perfect observation:

$$\forall t \geq 2, \ s_{k,t}^{\star} = \begin{vmatrix} \mathsf{narrowband} & \text{if } a_j(t-1) = \mathsf{narrowband}, j \in \{1, \ldots, K\} \\ \mathsf{wideband} & \text{otherwise.} \end{vmatrix} \tag{20}$$

with $a(1) = (\mathsf{narrowband}, \ldots, \mathsf{narrowband})$. By implementing these strategies, each player gets a utility which equals 3 whereas it was 1 in the static game version. Therefore, repeating the game and considering long-term utilities allows one to reach more efficient points at every stage of the game. This can be interpreted as a form of cooperation among the players. Thus far, we have mentioned two forms of cooperation namely, through bargaining and cooperative plans in repeated games. In Sec. IV, we will see that the coalition form offers another way of implementing cooperative solutions in games. From the above discussion, it follows that referring to strategic-form games as non-cooperative games and coalition games as cooperative games is questionable. Indeed, cooperation may exist in the former while players may still be selfish in the latter.

**Remark 2.** *In general, extensive-form games group all situations in which the players are allowed to have a sequential interaction, meaning that the move of each player is conditioned by the previous moves of all players in the game. This class of games is termed* dynamic games. *Repeated games are a subclass of dynamic games, in which the players face the same single-stage (static) game every period. Hence, while extensive-form games* *are not treated due to the lack of space needed to address their general aspect, repeated games, which represent a notable example, are included in this tutorial thanks to their broad field of application in the SP scenario.*

*Bayesian games:* When one wants to perform the direct maximization of a function while some of its parameters are unknown, a possible solution is to consider an expected version of the function of interest (e.g., think of the famous expectation-maximization algorithm). When solving a game, a similar approach can be adopted. In the presence of multiple decision makers, the problem is however more difficult. To understand this, assume that each player chooses a prior distribution over the parameters it does not know (e.g., the overall channel state): this is its *belief*. But, a player also has to assume what it knows about the belief of the other players. Going further, a player needs to have a belief about the belief on the other players on its own belief. This leads to the quite complex notion of hierarchy of beliefs. This approach seems to be inapplicable in practice. Why should an automaton or a computer implement such an elaborate level of reasoning? An important result of practical interest is that a simpler model might capture the whole hierarchy of beliefs. This model is known as Harsanyi's model [52] and it is very close in spirit to what is done in estimation problems in the presence of uncertain parameters. Once the game is formulated as a strategic-form (Bayesian) game, standard tools can be exploited. Although it is exactly an NE in the presence of expected utilities, an NE is called in this context a Bayesian equilibrium. Application examples of Bayesian games in the literature of SP for communications can be found in [53]. Therein, the unknown parameter is typically the communication channel state. In [54], the authors illustrate how Bayesian games are natural settings to analyze multiagent adaptive sensing systems.

## III. LEARNING EQUILIBRIA IN STRATEGIC-FORM GAMES

To better understand the relationship between the solution concepts described in Sec. II and algorithmic aspects, we will first consider some experiments, which were conducted by the biologist David Harper [55]. These experiments are of interest to better understand how equilibria can be achieved (learnt) by repeated interactions driven by simple decision-making rules. In winter 1979, Harper conducted experiments on a flock of 33 ducks on a lake in the botanic garden of Cambridge University, UK. Two observers who were acting as bread tossers were located at two fixed points on the lake surface 20 meters apart. The pieces of bread were thrown at regular intervals. For instance, one of the experiments assumes that the frequency of supply for one observer (called the least profitable site) is 12 items per minute whereas it was equal to 24 items per minute for the other observer. Fig. 8 represents the number of ducks at the least profitable site against time; the dots indicate the mean points while the vertical segments represent the dispersion of the measures. It is seen that after about a minute, the number of ducks at the least profitable site stabilizes around 11, which means that 22 ducks are at the most profitable site. The corresponding point is an NE: every duck which would switch to the other site in a unilateral manner would get less food. Fig. 8 shows that, at the beginning of the trial, each duck behaves like a conventional optimizer: most of the ducks goes to the most profitable site. This choice does not take into account that the site selection problem a duck faces with is not a conventional optimization problem but a game: what a duck gets does not only depend on its choice but also on others' choice. During the transient period, the ducks, which switch to the other site, realize they get more food at the least profitable site. Other ducks do so as long as an equilibrium is reached. Quite likely ducks do not know their utility functions and, more generally, the parameters of the game they play. They may hardly be qualified as rational players as well. Nonetheless, some sort of iterative "auction" process (known as tâtonnement) has led them to an NE showing that an NE can emerge as the result of repeated interactions between entities, which have only partial information on the problem and only implement primitive decision-making or learning rules. The main purpose of this section is precisely that of providing some learning rules (or SP algorithms) among many others from the vast literature of multi-agent learning, learning in games, or distributed optimization, which may lead to equilibria.

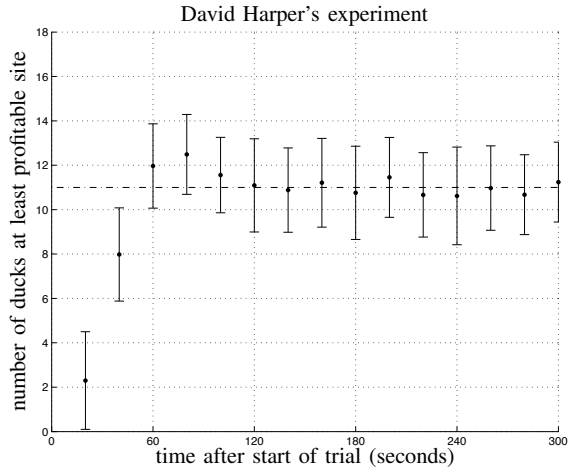

Fig. 8: When ducks are given the choice between two bread tossers for which the frequency of supply of the most profitable site is twice the least profitable, after switching a few times between the two sites, ducks stick to a given choice. The corresponding point is an NE.

*Remark.* Although in the remainder we only focus on distributed optimization and multi-agent learning algorithms as solution concepts for certain static game, it is worth observing that it may also be possible to interpret a multi-agent learning rule as a strategy of a certain dynamic game [22], showing also the existence of a relationship between learning and dynamic games.

### A. Best-response dynamics

BRD is a popular and simple learning rule, which may lead to equilibria. The BRD has been used in various disciplines but, as its use is specialized, the different instances of it are not always recognized as the same algorithm. Two instances of it are the Gauss-Seidel method [56] and the Lloyd-Max algorithm [57]. The Gauss-Seidel method is an iterative algorithm that allows to numerically solve a linear system of equations. Let us review this method in the special case of two unknowns $x_1, x_2$ and two observations $y_1, y_2$. The goal is to solve the system

$$\begin{pmatrix} a_{11} & a_{12} \\ a_{21} & a_{22} \end{pmatrix} \begin{pmatrix} x_1 \\ x_2 \end{pmatrix} = \begin{pmatrix} y_1 \\ y_2 \end{pmatrix} \tag{21}$$

where the entries $a_{kj}$ are assumed to be known and meet some classical conditions, which can be found in [56]. By denoting $(x_1(t), x_2(t))$ the value for the pair $(x_1, x_2)$ at iteration $t$, $x_1$ is updated as $x_1(t+1)$ which is obtained by solving $a_{11}x_1(t+1) + a_{12}x_2(t) - y_1 = 0$. Then, $x_2(t+1)$ is obtained by solving $a_{21}x_1(t+1) + a_{22}x_2(t+1) - y_2 = 0$. This can be interpreted as a game with two players in which $x_k$ is the action of player $k$ and setting (or making close) to zero $a_{kk}x_k + a_{k,-k}x_{-k} - y_k$ is its objective or cost function. The Gauss-Seidel method precisely implements the sequential BRD of the latter game.

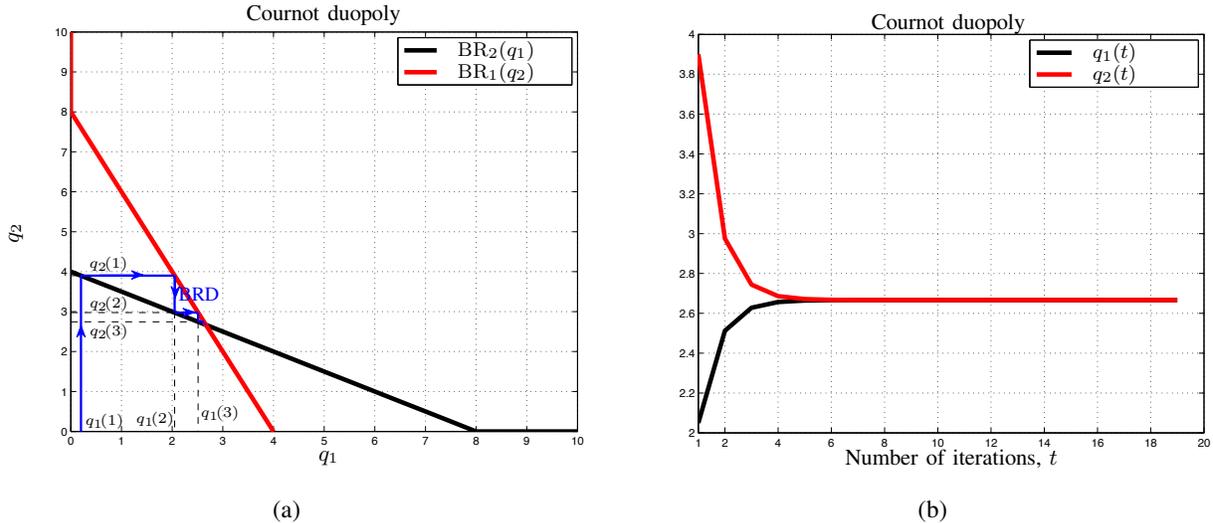

(a)                                                     (b)

Fig. 9: Illustration of the Cournot tâtonnement. This process, which is a special case of the sequential BRD algorithm, converges to the unique intersection point between the players' best-responses (i.e., the unique pure NE of the game). As well illustrated by the Cournot duopoly, convergence of sequential BRD is typically fast.

As observed in [58], another special instance of the BRD is the Lloyd-Max algorithm originally used for scalar quantization and extensively used nowadays in data compression techniques in information theory and SP. Designing a signal quantizer means choosing how to partition the source signal space into cells or regions and choosing a representative for each of them. It turns out that finding in a joint manner the set of regions and the set of representatives which minimize the distortion (namely, the quantization noise level) is a difficult problem in general. The Lloyd-Max algorithm is an iterative algorithm, in which each iteration comprises two steps. First, one fixes a set of regions and computes the best representatives in the sense of the distortion. Second, for these representatives, one updates the regions so that distortion is minimized. This procedure is repeated until convergence and corresponds to a special instance of the sequential BRD of a game with two players which have a common cost function. As seen in Sec. II-C, since the cost function is common, the game is potential. As explained a little further, convergence of the sequential BRD is guaranteed in such games.

**Example 4** (Cournot tâtonnement). *Another well-known instance of the BRD is the Cournot tâtonnement. It was originally introduced by Cournot in 1838 to study an economic competition between two firms where each one has to decide the quantity of goods to produce. In particular, Cournot showed that the following dynamical procedure converges: firm 1 chooses a certain quantity of goods $q_1(1)$, firm 2 observes the quantity produced by firm 1 and plays its best-response $q_2(2)$, i.e., the quantity maximizing its profit, firm 1 re-adjusts its quantity to this reaction to $q_1(3)$ in order for its benefit to be maximal and so forth. Cournot proved that after "a while"*

---

**Algorithm 1** BRD.

---

**set** $t = 0$

**initialize** $a_k(0) \in \mathcal{S}_k$ for all players $k \in \mathcal{K}$ (e.g., using a random initialization)

**repeat**

    **for** $k = 1$ to $K$ **do**

        **update** $a_k(t + 1)$ using (22) or (23)

    **end for**

    **update** $t = t + 1$

**until** $|a_k(t) - a_k(t - 1)| \leq \epsilon$ for all $k \in \mathcal{K}$

---

*this process converges to the so-called Cournot equilibrium, which can be shown to be the NE of the associated strategic-form game. This is what Fig. 9 illustrates. A possible application of the dynamical procedure above can be found in [59] in which the authors consider a competitive spectrum sharing scheme based on GT for a CR network consisting of a primary user and multiple secondary users sharing the same frequency spectrum. The spectrum sharing problem is modeled as an oligopoly market and a static game has been used to obtain the Nash equilibrium for the optimal allocated spectrum size for the secondary users.*

The BRD can be formulated for a game with an arbitrary number of players. In its most used form, the BRD operates in a sequential manner (sequential BRD) such that players update their actions in a round-robin manner. Within round $t + 1$ (with $t \geq 1$) the action chosen by player $k \in \mathcal{K}$ is computed as:[5]

$$a_k(t + 1) \in \mathrm{BR}_k\left[a_1(t + 1), \ldots, a_{k-1}(t + 1), a_{k+1}(t), \ldots, a_K(t)\right]. \tag{22}$$

An alternative version of the BRD operates in a simultaneous way meaning that all players update their actions simultaneously:

$$a_k(t + 1) \in \mathrm{BR}_k\left[a_{-k}(t)\right]. \tag{23}$$

The pseudocode of BRD for both instances is sketched in Algorithm 1. Observe that both can be applied to games in which the action sets are either continuous or discrete. If continuous, convergence means that the distance between two successive action profiles remains below a certain threshold $\epsilon > 0$. If discrete, convergence means that the action profile does not change at all (i.e., $\epsilon = 0$). When it converges, convergence points are typically pure NE (e.g., see [22]). There are no convergence results for general games using BRD. Most of the existing results rely on application-specific proofs. For example, [5] considers an application example of the BRD in SP for which an ad hoc proof for convergence is provided. However, if some special classes of games are considered, then there exist sufficient conditions under which the convergence of the sequential BRD to a pure NE is always guaranteed. For example, it is ensured when exact potential games and supermodular games are considered (see Sec. II-C and [22] for more details on this). In addition to this, the convergence of the sequential BRD is ensured when the best-responses are standard functions [60]. These results are summarized next.

---

[5]If there are more than one best actions, then one of them is chosen at random from the uniform probability distribution.

**Theorem 3** ([22]). *In potential and supermodular games, the sequential BRD converges to a pure NE with probability one.*

**Theorem 4** ([60]). *If the best-responses of a strategic-form game are standard functions, then the BRD converges to the unique pure NE with probability one.*

Unlike the sequential BRD, there does not seem to exist general results that guarantee the convergence of the simultaneous BRD. As shown in [61], a possible way out to ensure convergence is to let player $k$ update its action as $a_k(t+1) \in \overline{\mathrm{BR}}_k [a_{-k}(t)]$ where $\overline{\mathrm{BR}}_k [a_{-k}(t)]$ is defined as

$$\overline{\mathrm{BR}}_k [a_{-k}(t)] = \arg \max_{a_k \in \mathcal{A}_k} u_k(a_k, a_{-k}(t)) + \kappa \|a_k - a_k(t)\|^2 \tag{24}$$

with $\kappa \geq 0$. The term $\|a_k - a_k(t)\|^2$ acts as a stabilizing term, which has a conservative effect. If $\kappa$ is large, this term is minimized by keeping the same action. By choosing $\kappa$ in an appropriate manner, [61] shows that the simultaneous BRD associated with the modified utility converges.

Now we consider an application example that will be developed throughout this section to illustrate the different algorithms and notions under consideration. In particular, it allows us to extract sufficient conditions under which the sequential BRD converges.

**Example 5** (Power allocation games in multi-band interference channels). *Consider a wireless communication system, which comprises $K$ transmitter-receiver pairs. Each transmitter wants to communicate with its own receiver. More precisely, transmitter $k \in \{1, \ldots, K\}$ (player $k$) has to allocate its available power (denoted by $P$) among $N$ orthogonal channels or frequency bands to maximize its own transmission rate $u_k = \sum_{n=1}^{N} \log_2 (1 + \gamma_{k,n})$ where $\gamma_{k,n}$ is the signal-to-interference-plus-noise ratio (SINR) at receiver $k$ over band $n$, which is defined as*

$$\gamma_{k,n} = \frac{h_{kk,n} p_{k,n}}{\sigma^2 + \sum_{\ell \neq k} h_{\ell k,n} p_{\ell,n}} \tag{25}$$

*where $p_{k,n}$ is the power transmitter $k$ allocates to band $n$, $h_{\ell k,n} \geq 0$ is the channel gain associated with the link from transmitter $\ell$ to receiver $k$ over band $n$, and $\sigma^2$ accounts for the thermal noise. Denote by $p_k = (p_{k,1}, \ldots, p_{k,N})$ the power allocation vector of transmitter $k$. Two scenarios in terms of action space are considered:*

$$\mathcal{A}_k^{\mathrm{PA}} = \left\{ p_k \in \mathbb{R}_+^N : \sum_{n=1}^{N} p_{k,n} \leq P \right\} \text{ and } \mathcal{A}_k^{\mathrm{BS}} = \{ P e_1, \ldots, P e_N \} \tag{26}$$

*where PA stands for power allocation and BS for band selection, and $e_1, \ldots, e_N$ represents the canonical basis of $\mathbb{R}^N$ (i.e., $e_1 = (1, 0, \ldots, 0), e_2 = (0, 1, 0, \ldots, 0)$ and so on). The two corresponding strategic-form game will be denoted by $\mathcal{G}^{\mathrm{PA}}$ and $\mathcal{G}^{\mathrm{BS}}$.*

A sufficient condition for the sequential BRD to converge for the game $\mathcal{G}^{\mathrm{PA}}$ has been provided in [62]. The condition is that the spectral radius $\rho$ of certain matrices $\boldsymbol{H}(j)$ are strictly less than one:

$$\forall j \in \mathcal{K}, \rho(\boldsymbol{H}(j)) < 1 \text{ with } H_{k\ell}(j) = \begin{vmatrix} 0 & \text{if } k = \ell \\ \frac{h_{\ell j}}{h_{kj}} & \text{if } k \neq \ell. \end{vmatrix} \tag{27}$$

Condition (27) is useful for the general case of the multi-band interference channel and roughly means that the interference level on every band should not be too high. However, as shown in [48], the sufficient condition holds with probability zero (randomness stems from the fact that the channel gains $h_{k\ell,n}$ are assumed to be realizations of a *continuous* random variable) in the special case of the multi-band multiple access channel, which corresponds to have only one receiver of interest for all the transmitters. In the latter case, the SINR takes a more particular form, which is

$$\gamma_{k,n} = \frac{h_{k,n}p_{k,n}}{\sigma^2 + \sum_{\ell \neq k} h_{\ell,n}p_{\ell,n}} \tag{28}$$

where $h_{k,n}$ is the channel gain of the link between transmitter $k$ and the receiver for band $n$. Remarkably, in this particular setting $\mathcal{G}^{\mathrm{PA}}$ and $\mathcal{G}^{\mathrm{BS}}$ can be shown to be exact potential games [48] with potential function

$$\Phi = \sum_{n=1}^{N} \log_2 \left( \sigma^2 + \sum_{k=1}^{K} h_{k,n}p_{k,n} \right). \tag{29}$$

Exact potentiality of games guarantees the convergence of the sequential BRD to a pure NE. In game $\mathcal{G}^{\mathrm{PA}}$, the sequential BRD consists in updating the power level according to a water-filling formula:

$$p_{k,n}(t+1) = \left[ \frac{1}{\omega_k} - \frac{p_{k,n}(t)}{\gamma_{k,n}(t)} \right]^{+} \tag{30}$$

where $[x]^{+} = \max(0, x)$, $\omega_k$ is the Lagrangian multiplier associated with the inequality constraint $\sum_{n=1}^{N} p_{k,n} \leq P$, and $\gamma_{k,n}(t)$ is the SINR at receiver $k$ over band $n$ at time $t$. The solution is known as iterative water-filling algorithm (IWFA) and was introduced for the multi-band interference channel in [63]. In its most general form, the sequential BRD algorithm in (22) is quite demanding in terms of *observation* since each player has to observe the actions played by the others. In the case of the IWFA, it is seen that that only knowledge of the SINR $\gamma_{k,n}(t)$ is required to implement the BRD, which is basically an aggregate version of the played actions: this information can easily be estimated at the receiver and fed back to player $k$ for updating its transmit power. When converging, the IWFA, and more generally the sequential BRD, does it quite *fast*: convergence is typically observed after a few iterations [48]. Intuitively, the feature of fast convergence stems from the fact that the BRD relies on a detailed knowledge of the problem at hand. Typically, the utility functions are assumed to be known. When this knowledge is not available, instead of considering *highly structured* distributed optimization algorithms such as the BRD, one may consider multi-agent learning algorithms, which are typically much less demanding in terms of modeling the problem, as discussed in the next subsections. However, before moving to such techniques, an alternative version of the BRD is considered, which operates on probability distributions over actions (instead of pure actions) and is referred to as the fictitious play (FP) algorithm. Considering the FP algorithm allows us to better understand the iterative structure of many learning algorithms, particularly the one considered in Sec. III-B.

The original version of the FP algorithm assumes discrete action sets, which is what is also assumed next. It should be stressed that the BRD is generally not well suited to the discrete case. For example, when applied to $\mathcal{G}^{\mathrm{BS}}$ it converges in the scenario of multi-band multiple access channels while it does not converge in the multi-band interference channel case as cycles appear [64]. This is quite frequent in games with discrete actions. Therefore, learning algorithms such as the one described in Sec. III-B are not only useful to assume less structure on the problem but also to deal with the *discrete* case. From now on, we thus assume that

$$\mathcal{A}_k = \{a_{k,1}, \ldots, a_{k,N_k}\} \tag{31}$$

where $|\mathcal{A}_k| < +\infty$. The FP algorithm, introduced by Brown in 1951 [65], is a BRD algorithm in which empirical frequencies are used. Working with probability distributions is very convenient mathematically. Although mixed strategies are exploited, this does not mean that mixed NE are sought. In fact, pure NE can be shown to be attracting points for all the dynamics, which are considered in this survey. This means that, under appropriate conditions, mixed strategies tends to pure strategies as the number of iterations grows large. The empirical frequency of use of action $a_k \in \mathcal{A}_k$ for player $k \in \mathcal{K}$ at time $t+1$ is defined by:

$$\pi_{k,a_k}(t+1) = \frac{1}{t+1}\sum_{t'=1}^{t+1}\mathbb{1}_{\{a_{k,t'}=a_k\}} \tag{32}$$

where $\mathbb{1}$ is the indicator function. If player $k$ knows $\pi_{-k,a_{-k}}(t)$ (i.e., the empirical frequency of use of the action profile $a_{-k}$ at time $t$), then it can compute its own expected utility and eventually choose the action maximizing it. Observe that the computation of $\pi_{-k,a_{-k}}(t)$ requires to observe the actions played by the others. As for BRD, this knowledge can be acquired only through an exchange of information among the players.[6]

In its simultaneous form, the FP algorithm operates as follows:

$$a_k(t+1) \in \arg\max_{a_k \in \mathcal{A}_k} \sum_{k=1}^{K} \pi_{-k,a_{-k}}(t)u_k(a_k, a_{-k}). \tag{33}$$

The important point we want to make about the FP algorithm is about the structure of the empirical frequencies. As a matter of fact, they can be computed in a recursive fashion as:

$$\pi_{k,a_k}(t+1) = \frac{1}{t+1}\sum_{t'=1}^{t+1}\mathbb{1}_{\{a_{k,t'}=a_k\}} = \frac{1}{t+1}\sum_{t'=1}^{t}\mathbb{1}_{\{a_{k,t'}=a_k\}} + \frac{1}{t+1}\mathbb{1}_{\{a_{k,t+1}=a_k\}}$$

$$= \pi_{k,a_k}(t) + \lambda_k^{\text{FP}}(t)\left[\mathbb{1}_{\{a_{k,t+1}=a_k\}} - \pi_{k,a_k}(t)\right] \tag{34}$$

with $\lambda_k^{\text{FP}}(t) = 1/(t+1)$ The last line translates the fact that the empirical frequency at time $t+1$ can be computed from its value at time $t$ and the knowledge of the current action. More interestingly, it emphasizes a quite general structure which is encountered with many iterative and reinforcement learning (RL) algorithms, as seen in the remainder of this section.

## B. Reinforcement learning

Originally, RL was studied in the context of single-player (or single-automaton) environments with a *finite* set of actions. A player receives a numerical utility signal and updates its strategy. The environment provides this signal as a feedback for the sequence of actions that has be taken by the player. Typically, the latter relates the utility signal to actions previously taken in order to learn a mixed strategy which performs well in terms of average utility. In a multi-player setting, RL is inherently more complex since the learning process itself changes the thing to be learned. The main objective of this subsection is to show that feeding back to the players only the realizations of their utilities is enough to drive seemingly complex interactions to a steady state or, at least, to a predictable evolution of the state. In RL algorithms, players use their experience to choose or avoid certain actions based on their consequences. Actions that led to satisfactory outcomes will tend to be repeated in the future, whereas actions that led to unsatisfactory experiences will be avoided. One of the first RL algorithms was proposed by Bush and Mosteller in [66] wherein each player's strategy is defined by the probability of undertaking each of the available actions. After every player has selected an action according to its probability, every player receives the corresponding utility and revises the probability of undertaking that action according to a reinforcement policy. More formally, let $u_k(t)$ be the value of the utility function of player $k$ at time $t$, and denote by $\pi_{k,a_{k,n}}(t)$ the probability player $k$ assigns to action $a_{k,n}$ at time $t$. Then, the Bush and Mosteller RL algorithm operates as follows:

$$\pi_{k,a_{k,n}}(t+1) = \pi_{k,a_{k,n}}(t) + \lambda_k^{\text{RL}}(t) u_k(t) \left[ \mathbb{1}_{\{a_k(t)=a_{k,n}\}} - \pi_{k,a_{k,n}}(t) \right] \tag{35}$$

where $0 < \lambda_k^{\text{RL}}(t) < 1$ is a known function that regulates the learning rate of player $k$ (it plays the same role as the step-size in the gradient method). As seen, the updating rule given by (35) has the same form of (34) but one of the strengths of the algorithm corresponding to (35) is that each player only needs to observe the *realization* of its utility function and nothing else. It can therefore be applied to any finite game. Convergence is ensured for classes of games such as potential games and supermodular games, introduced in Sec. II-C.[7] As for the BRD, convergence points are either pure NE or boundary points. The price to be paid for the high flexibility regarding the environment and the absence of strong assumptions on its structure is that the RL algorithm in (35) usually requires a *large* number of iterations to converge compared to the BRD algorithm.

---

[6]For example, in the two-player CR's dilemma (Example 2), if CR1 has knowledge of the number of times that CR2 has picked narrowband or wideband up to time $t$, then CR1 can easily compute $\pi_{2,a_2}(t)$ through (32).

All the above distributed algorithms (namely, the BRD algorithm, the FP algorithm, and the considered RL algorithm) are attractive since they only rely on partial knowledge of the problem. On the other hand, convergence points are typically pure NE, which in most cases are inefficient. Often, points which Pareto-dominate the NE points can be shown to exist. A nontrivial problem is how to reach one of them in a distributed manner. We will not address this challenging task in this survey. Rather, we will provide one learning algorithm that allows players to reach a CCE. This may be more efficient than a pure or mixed NE, since the latter is a special instance of it.

### C. Regret matching learning algorithm

The key auxiliary notion, which is exploited for regret matching (RM) learning algorithms is the notion of *regret* [67], which is eventually exploited to assign a certain probability to a given action. The regret player $k$ associates with action $a_{k,n}$ is the difference between the average utility the player would have obtained by always playing the same action $a_{k,n}$, and the average utility actually achieved with the current strategy. Mathematically, the regret at time $t$ for player $k$ is computed as

$$\forall n \in \{1, \ldots, N_k\}, \ \ r_{k,a_{k,n}}(t+1) = \frac{1}{t} \sum_{t'=1}^{t} u_k(a_{k,n}, a_{-k}(t')) - u_k(a_k(t'), a_{-k}(t')). \tag{36}$$

RM relies on the assumptions that at every iteration $t$, player $k$ is able to evaluate its own utility – i.e., to calculate $u_k(a_k(t), a_{-k}(t))$ – and to compute the utility it would have obtained if it had played another action $a'_k$ (i.e., $u_k(a'_k, a_{-k}(t))$). In [67], the rule for updating the probability player $k$ assigns to action $a_{k,n}$ is as follows:

$$\pi_{k,a_{k,n}}(t+1) = \frac{\left[ r_{k,a_{k,n}}(t+1) \right]^+}{\sum_{n'=1}^{N_k} \left[ r_{k,a_{k,n'}}(t+1) \right]^+}. \tag{37}$$

---

[7]The convergence of RL algorithms is also ensured for dominance solvable games [22], which are not treated in this survey due to space limitations.

---

**Algorithm 2** Regret matching learning algorithm

---

**set** $t = 0$

**initialize** $\pi_k(0)$ s.t. $\sum_{n=1}^{N_k} \pi_{k,n}(0) = 1$ for all players $k \in \mathcal{K}$ (e.g., using a random initialization)

**repeat**

    **for** $k = 1$ to $K$ **do**

        **for** $n = 1$ to $N_k$ **do**

            **update** $r_{k,n}(t+1)$ using (36)

        **end for**

        **for** $n = 1$ to $N_k$ **do**

            **update** $\pi_{k,n}(t+1)$ using (37)

        **end for**

        **choose** $a_k(t+1)$ according to the distribution $\pi_k(t+1)$

    **end for**

    **update** $t = t+1$

**until** $|a_k(t) - a_k(t-1)| \leq \epsilon$ for all $k \in \mathcal{K}$

---

Table II: Main features for the BRD, RL, and RM algorithms.

| | **BRD** | **RL** | **RM** |
|---|---|---|---|
| *Action sets* | continuous or discrete | discrete | discrete |
| *Convergence* | sufficient conditions | sufficient conditions | always guaranteed |
| *Convergence points* | pure NE or boundary points | pure NE or boundary points | CCE |
| *Convergence speed* | fast | slow | medium |
| *Efficiency of convergence points* | typically low | typically low | typically medium |
| *Observation typically required* | actions of the others | value of the utility function | actions of the others |
| *Knowledge typically required* | utility functions and action sets | action sets | utility functions and action sets |

If, at time $t+1$, player $k$ has a positive regret for every action, it implies that it would have obtained a higher utility by playing the same action during the whole game up to iteration $t+1$, instead of playing according to the distribution $\pi_k(t) = (\pi_{k,a_{k,1}}, \ldots, \pi_{k,a_{k,N_k}})$. The updating rule (37) has a very attractive property: it is with *no regret* [67]. The consequence of this property is expressed through the following result.

**Theorem 5** (Convergence of the RM algorithm)**.** *In any finite game, when updated as* (37)*, the empirical frequencies of the action profile always converges almost surely to the set of CCE.*

Observe that in those games wherein CCE, CE, mixed NE, and pure NE coincide (such as for example in the simple CR's dilemma introduced in Example 2), then a unique CCE exists, which is a pure NE. In this particular setting, RM does not provide any performance gain over the BRD. However, in most cases the RM algorithm has the potential to perform better than distributed algorithms such as the BRD. This is what is illustrated in Sec. III-D. In the CR context, an application example supporting this statement can be found in [68].

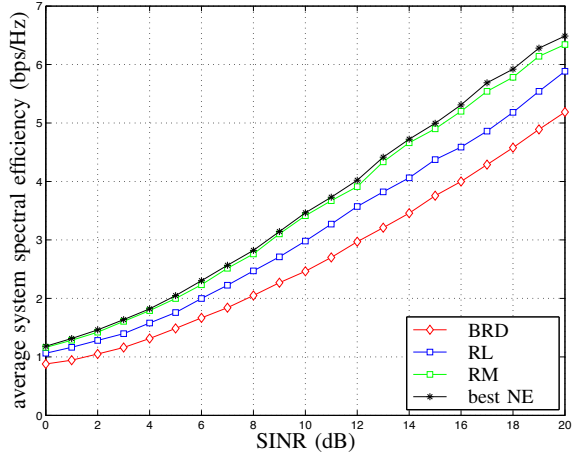

Fig. 10: RM has always the potential to perform better than BRD since pure NE are special cases of CCE. The figure shows that this is effectively the case for the sum-rate of the considered distributed power allocation problem under the given simulation setup [64].

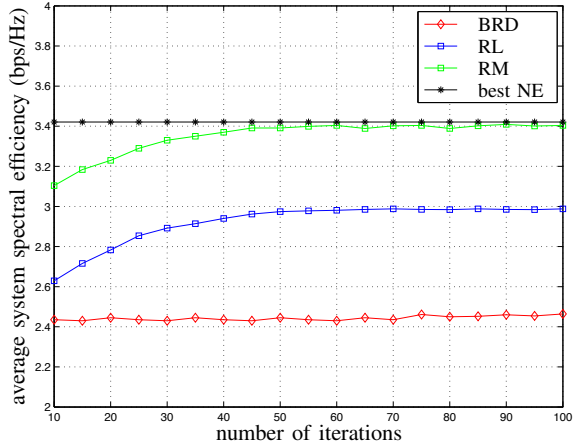

Fig. 11: Average system spectral efficiency as a function of the number of iterations at a fixed SINR of $10\,\text{dB}$.

### D. Illustration and comparison analysis

Table II summarizes the different features of the three classes of distributed algorithms, which have been discussed throughout this section. Here, we consider a special instance of game $\mathcal{G}^{\text{BS}}$ in which only two transmitters and two receivers are operating and two bands are assumed, and each transmitter has to select one single band [69]. Fig. 10 depicts the performance in terms of sum-utility (i.e., the transmission sum-rate) as a function of the SINR for both BRD and RM algorithms. As shown in Fig. 10, the RM learning algorithm is more efficient in terms of sum-rate than the BRD algorithm. In fact, here the performance of the CCE, which is obtained by implementing the RM learning algorithm, is very close to the performance of the best pure NE of the game. On the other hand, the BRD is observed to converge to a pure NE, which does not coincide with the best NE. Although this is not what is observed generically, there may exist some initial points for which the BRD performs better than the RM algorithm. This raises a challenging problem that is to characterize the relation between the initial and convergence points, which is a challenging and open issue. Note that if the RL algorithm is considered, the same issue would appear. The performance of the RL algorithm for the special case of interest would also strongly depend on the initial point. The main drawback of using the RL algorithm would be the number of iterations needed for convergence (when the algorithm effectively converges), as shown in Fig. 11.

*E. Consensus algorithms*

One last type of algorithms described in this section accounts for consensus algorithms. These algorithms rely on a *strong coordination* between the players. This is achieved at the price of a quite strong observation assumption: the corresponding updating rule requires explicit knowledge of the actions chosen by the other agents or players. As a result of this assumption, an *efficient* solution can be attained at convergence. For instance, assume that the players' actions are real numbers, $\forall k \in \mathcal{K}, a_k \in \mathbb{R}$, and assume that the network should be designed to operate at a given point $a^\star = (a_1^\star, \ldots, a_K^\star) \in \mathbb{R}^K$ referred to as *consensus*. This point must be attained by each player through a certain iterative and distributed procedure involving exchanges among the agents; of course reaching a point which is globally efficient may not be possible. A simple instance of a consensus algorithm (see e.g., [70]) is as follows:

$$a_k(t+1) = a_k(t) + \sum_{j \in \mathcal{A}_k} \beta_{k,j} \left( a_j(t) - a_k(t) \right) \tag{38}$$

where $t$ is the iteration index, $\mathcal{A}_k$ represents the neighbors of agent $k$, and $\beta_{k,j}$ is some weight that player $k$ places on the action or state of player $j$. Simple sufficient conditions can be stated under which such an algorithm converges [70]. Indeed, the convergence analysis amounts to studying the dynamical system $a(t+1) = \mathbf{C}a(t)$ where the matrix $\mathbf{C}$ follows from (38). The convergence properties of consensus algorithms have been studied under several interaction models (delays in information exchange, connectivity issues, varying topologies and noisy measurements) and can be usually ensured by construction of the algorithm itself. However, this requires a well determined topology for the network and also a quite large amount of information to be exchanged, especially in comparison with the other learning algorithms described above. Surprisingly, there has been relatively little research that explicitly links consensus problems or, more generally cooperative control problems, to the very relevant branches of learning in game literature or multi-agent system literature that address coordination problems. Most of the attempts in this context aim at establishing a connection between coordination problems and potential games [51], [71]. To conclude this section, a simple application of consensus is given below.

**Example 6** (Detection with sensor networks)**.** *Consider a wireless sensor network in which each sensor can only communicate with the sensors within its transmitted signal range. Each sensor has to decide whether a tectonic plate is active or not (e.g., to detect earthquakes). The action to be taken by each sensor is assumed to be binary* **active** *or* **not active***. To decide whether a plate is active or not by using all the measurements and associated decisions, a consensus algorithm as that in* (38) *can be implemented [70].*

## IV. Coalition-form games

As discussed in Sec. II, strategic-form games mainly focus on the strategic choices of individual players and on what strategies each player would choose to reach its goal. More importantly, strategic-form representations often deal with noncooperative cases in which players are assumed to act selfishly, individually, and without any side payment, cooperation, or exchange of communication. In contrast, many SP applications require some sort of cooperation between the players. For example, it is more and more common to form virtual arrays of antennas, sensors, or telescopes to improve estimation or detection accuracy; this type of operations requires communication and partial-to-full cooperation between the players. Cooperative networking, in which devices can, for example, cooperatively route their packets at the network layer, is also a typical application where cooperation is needed. In such cases, given the cooperative nature of the system, players may form groups among one another, in an effort to improve their state and position in the game. Thus, we now deal with groups of players or *coalitions* that act in a coordinated manner. Inside each such coalition, the players may still be choosing strategies, similar to a strategic-form game, but overall, the goal in here is to analyze the formation of the coalitions given the possibility of communication between the players.

*Coalition-form games* provide an appropriate representation for such situations in which groups or coalitions (subsets) of players can work together in a game. In such games, we are typically concerned about the options available to coalitions, the possible coalitions that will form, and how the utility received by the coalition as a whole can be divided among its members in a way to sustain cooperation. This amounts to assuming the existence of a mechanism which imposes a particular action or, more in general, series of actions on each player. This mechanism can for example result from a binding agreement among the players or from a rule imposed by a designer.

The coalition form is suitable to model a number of problems. On the one hand, it is the only game-theoretic tool available to predict and characterize how groups of players can weight and evaluate the mutual benefits and cost from cooperation and, then, decide on whether or not to work together and form binding agreements. On the other hand, when the coalition form is found to be suitable to model the problem at hand, one of its strengths is that it may lead to a solution, which is more efficient than in the case in which no coordination occurs. Moreover, the coalition form provides a suite of tools that allow us to evaluate fairness, stability, and efficiency, when players in a game are able to coordinate and communicate with each other, prior to making decisions.

## A. Coalition-form games and bargaining theory

One important distinction to make is that between NBS (more generally, the bargaining theory) and coalition-form games. In the game-theoretic literature [23], both Nash bargaining and coalition-form games are often grouped under the umbrella of cooperative games. This classification mainly stems from the fact that, in both cases, the players may coordinate their strategies and are, in general, cooperative. However, the NBS is restricted to the scenario in which two or more players want to share a resource, and they are, a priori, willing to cooperate in this resource sharing, provided that the "terms" of cooperation are fair. Then, the question becomes the following: given the players' initial positions (which is generally the max-min or NE solution using their individual utility functions), which have to be feasible, how should they split the rest of the resource being shared? Subsequently, as detailed in Sec. II-B, the NBS follows an axiomatic approach. In this regard, the NBS provides a *unique* allocation that answers this question.

Now, although the original solution proposed by Nash was restricted to two players, the idea of Nash bargaining has then been extended to the general multiple-player game. This extension has been particularly popular in the SP community, where the analogies between Nash bargaining and the famous proportional fair resource allocation mechanisms have been drawn and exploited. Important examples include consensus algorithms, resource allocation, and optimal beamforming [4], [9], [39]. Nonetheless, even with this extension, the overall Nash bargaining problem remains the same – how to share a resource between *all* players, so as to: *i*) satisfy the Nash bargaining axiom and *ii*) improve the players' overall utility.

In contrast, coalition-form games address a different problem: how cooperative coalitions among different players can be formed given the mutual benefits and costs for cooperation. Therefore, coalition-form games are not restricted to a fair resource-sharing problem such as in the NBS. In contrast, they investigate a much more generic problem. Coalition-form games study how to stabilize and maintain cooperative coalitions between groups of players, in any situation, not just resource allocation. In contrast, for a bargaining problem, it is assumed that: *i*) all players are willing to cooperate, *ii*) there is no cost for cooperation, and *iii*) the cooperation is reduced to share a resource.

Therefore, in terms of objectives, the two approaches are different. However, the NBS can be used as an axiomatic solution for distributing the utility inside a "formed" coalition, in a fair manner (in the Nash bargaining sense). However, even though the bargaining solution will satisfy the NBS fairness axioms, it will not necessarily stabilize the coalition, in the sense that some players may still want to leave this coalition and form other coalitions, if the NBS is used to distribute the utilities. Thus, to study large-scale cooperation and coalition-formation processes, one must use solution concepts and algorithms that are much more general than the NBS. This motivates the need for coalition-form games.

*B. Coalition-form game models*

In this section, we use the notation $\mathcal{C}$ to refer to a given subset of the set of players $\mathcal{K} = \{1, ..., K\}$. The notation $2^{\mathcal{K}}$ is used to denote the power set associated with $\mathcal{K}$. For example, if $\mathcal{K} = \{1, 2\}$, then $2^{\mathcal{K}} = \{\emptyset, \{1\}, \{2\}, \{1, 2\}\}$. A coalition game is defined by the pair $(\mathcal{K}, v)$, where $v$ is the value of a coalition that is a function or mapping that provides a characterization of the utility (or utilities) achieved by the players that belong to a certain coalition.

In essence, for classical models of coalition games, depending on the definition of $v$, we can distinguish between non-transferable-utility (NTU) games and transferable utility transferable-utility (TU) games:

- NTU games: coalition actions result in utilities to individual coalition members;
- TU games: utilities are given to the coalition and then divided among its members.

In an NTU game, the formation of a coalition $\mathcal{C} \subseteq \mathcal{K}$ leads to changes of the individual utilities of the players within $\mathcal{C}$; however, there is no single value that can be used to describe the overall coalition utility. In contrast, in a TU setting, a single-valued function can be used to determine the overall utility of a coalition. Subsequently, the individual utilities can be viewed as a sharing of this single-valued gain.

NTU and TU games can be further categorized into characteristic-function (CF) games or partition-function (PF) games:

- PF games: the utility of a coalition $\mathcal{C} \subseteq \mathcal{K}$ depends on the actions chosen by the other coalitions in $\mathcal{K} \setminus \mathcal{C}$;

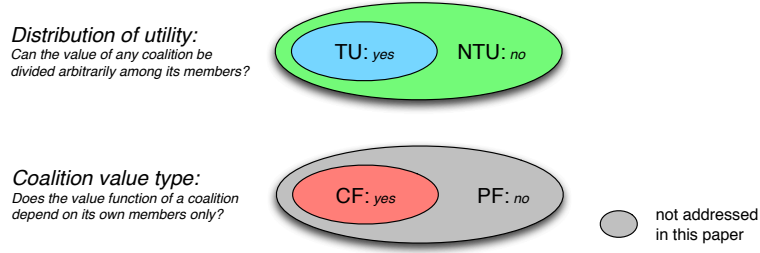

Fig. 12: Classification of coalition-form games.

• CF games: the utility of a coalition $\mathcal{C}$ only depends on the action chosen by the members of $\mathcal{C}$.

Both CF and PF games admit many applications in SP. The latter is particularly useful for cases in which externalities, such as interference or delay in communication networks, are present and depend on the coalition actions of the players. For simplicity, our focus will be on CF games. This classification of coalition-form games is shown in Fig. 12, emphasizing the fact the TU and CF game are special classes of NTU and PF games, respectively.

*1) NTU games:* The formal definition of a coalition-form NTU game with characteristic function often follows the form introduced by Aumann and Peleg in [72], which states that:

**Definition 13** (NTU coalitional games with CF). *An NTU game with CF is given by a pair* $(\mathcal{K}, v)$: $\mathcal{K} = \{1, ..., K\}$ *is called the set of players and $v$ is the characteristic function. The latter is a set-valued function*

$$
\begin{aligned}
v: \quad 2^{\mathcal{K}} \quad &\rightarrow \quad \mathbb{R}^{\mathcal{K}} \\
\mathcal{C} \quad &\mapsto \quad v(\mathcal{C})
\end{aligned}
\tag{39}
$$

*such that for every coalition $\mathcal{C} \subseteq \mathcal{K}$, $v(\mathcal{C})$ is a closed convex subset of $\mathbb{R}^{\mathcal{K}}$ that contains the utility vectors that players in $\mathcal{C}$ can achieve.*

In other words, in an NTU game, the value is a set of payoff vectors that can be achieved by the players in the game. A coalition game is therefore said to be NTU if the value or utility of a coalition cannot be arbitrarily apportioned between the coalition's members. For an NTU model, the players do not value a given coalition in the same way. Instead, for every coalition, one or more vectors of individual payoffs will be achieved. For example, when investigating a bargaining situation in which players cannot share their utilities, we can view the NBS vector as an example of an NTU allocation. In SP problems, casting a problem as an NTU coalition game strongly depends on the metrics being optimized. Some metrics such as energy are individual and thus NTU by design, while others (such as for example the sum-rate) are not necessarily NTU.

*2) TU games:* A special case of NTU games is given by TU games. In TU games, $v(\mathcal{C})$ is a real value that represents the total utility obtained by the coalition $\mathcal{C}$. This is what the following model translates.

**Definition 14** (TU coalitional games with CF). *A TU game with CF is given by a pair* $(\mathcal{K}, v)$: $\mathcal{K} = \{1, ..., K\}$

*is called the set of players and $v$ is the characteristic function. The latter is given by:*

$$v: \quad 2^{\mathcal{K}} \quad \to \quad \mathbb{R}$$
$$\mathcal{C} \quad \mapsto \quad v(\mathcal{C}). \tag{40}$$

The TU property means that this worth can be divided in any manner among the coalition members. The values in TU games are thought of as monetary values that the members in a coalition can distribute among themselves using an appropriate rule (one such rule being an equal distribution of the utility). In SP problems, one typical example in which the TU property is applicable is the case in which groups of devices aim to optimize a certain sum-rate. Given that a sum-rate can virtually be divided between the devices via a proper choice of transmit signal (or, more specifically, a power allocation), we can view the sum-rate as a TU metric.

**Remark 3.** *In practice, we can convert an NTU game to a TU game for the purpose of analysis. One way to do so is to define the TU value function as being the sum of the individual payoffs of the players. Even though the actual division of this sum cannot be done in this case in an arbitrary manner, we can still use the TU model to understand how the system would behave under cooperation. In this case, we can consider this single-valued TU utility as being a total revenue achieved by the entire utility, with the individual divisions being the virtual monetary gain that is provided to each player, if those players are to act within a coalition.*

*3) Canonical game:* For any type of coalition-form game, the primary goal is to develop strategic algorithms and mechanisms that allow to characterize and predict which coalitions will form, when, and how. Given this goal, we often refer to coalition games as *coalition-formation games*. However, one special case occurs when the value of a coalition is non-decreasing with respect to the size of the coalition. Here, cooperation is always beneficial and the costs of cooperation are negligible. In this specific case, the game is said to be *superadditive*, which is formally defined as follows for the TU case:

$$v(\mathcal{C}_1 \cup \mathcal{C}_2) \geq v(\mathcal{C}_1) + v(\mathcal{C}_2). \tag{41}$$

In this setting, it is trivial to see that the grand coalition of all players will yield the maximum utility. However, this does not mean that this grand coalition will always form. In fact, unless the total gains are properly distributed to the grand coalition's members, some of those members may deviate and form their own coalitions. In such scenarios, the coalition-formation game is simply reduced to the so-called *canonical game model*, in which the goal is no longer to form coalitions, but rather to study ways in which the grand coalition of all players can be sustained. This will lead to many solutions that look at fairness and stability, as detailed in the next subsection.

**Remark 4.** *This basic definition of the various coalition-form game types can be used as a basis to develop more advanced model. For example, if a player may belong simultaneously to multiple coalitions, we can define the framework* overlapping coalition-formation (OCF) games. *In SP, this could be used to model applications such as sharing of sensor data between multiple cooperating groups. In OCF scenarios, one must redefine the way a coalition-form game is presented. One approach is to represent a coalition by a $1 \times K$ vector $r$ whose element $r_i$ represents the amount of resources that player $i$ has shared with this coalition. For such OCF scenarios, notions of stability or fairness must now be extended to the new representation and definition of a coalition.*

Given this overview on how to represent a coalition-form game, our next step is to discuss the solution concepts and main results.

### C. Solution concepts

For coalition-form games, we can distinguish two features for the solution: *stability* and *fairness*. On the one hand, the solution of coalition-formation game must ensure that the formed coalitions are not susceptible to deviations by individual members or even sub-groups of members. On the other hand, given that coalition formation entails a division of utility, a suitable coalition-form solution must ensure fairness when dividing or allocating the various utilities. Balancing the two goals of fairness and stability is challenging and strongly dependent on factors such as the structure of the value function, the goals of the players, and the application being studied.

The solution of a coalition-form game can further be classified into two additional types: *set-valued solutions* and *single-valued solutions*. Set-valued solutions refer to solutions that can guarantee stability or fairness via more than one cooperative strategy. How to choose the most appropriate point within a set-valued solution becomes an important problem. This is reminiscent of the multiplicity of NE in strategic-form games. In contrast, single-valued solutions provide a unique strategy which achieves a given fairness or stability criteria. Practically, although both set-valued and single-valued solutions can be used for both fairness and stability, most existing set-valued solutions are focused on stability while single-valued solutions are tailored towards fairness.

While both solutions can apply to any type of coalition-form game, for ease of exposition, in this section we restrict our attention to CF games that are superadditive and TU. By doing so, the overall solution can be viewed as a distribution of utilities that can maintain the stability and fairness within the grand coalition. Nonetheless, throughout our discussions, we will point out the key aspects needed to extend the solutions to the more general coalition-formation cases. Moreover, in Sec. V, we will discuss algorithmic implementations that can provide more insights on solving coalition-formation games.

*1) The core:* The most popular set-valued solution of a coalition-form game is the *core* [73]. The core is the set of payoff allocations which guarantees that no group of players has an incentive to leave the grand coalition $\mathcal{K}$ to form any other coalition $\mathcal{C} \subset \mathcal{K}$. For a TU game, we let $x$ be the $1 \times K$ vector of individual user utilities. Here, we must have group rationality, i.e., $\sum_{i \in \mathcal{K}} x_i = v(\mathcal{K})$. In other words, the total allocation must sum to the entire value of the grand coalition. In addition, we define a payoff vector $x$ to be *individually rational* if every $x_i \geq v(\{i\}), \forall i$. This implies that an individually rational payoff vector ensures that no player will obtain a lower payoff by joining the grand coalition. Consequently, *the core* of a coalition game is defined as the set $\mathcal{S}$ of individually rational and group rational payoff vectors as follows:

$$\mathcal{S} = \left\{ x : \sum_{i \in \mathcal{K}} x_i = v(\mathcal{K}) \text{ and } \sum_{i \in \mathcal{C}} x_i \geq v(\mathcal{C}) \ \forall \ \mathcal{C} \subseteq \mathcal{K} \right\}. \tag{42}$$

In simple terms, the core of a coalition game is the set of payoff allocations that ensure that no group of players would have an incentive to leave the grand coalition and form their own individual coalition. The core guarantees stability with respect to any deviation by any group of players. However, even though the core guarantees stability and, for the superadditive case, we can easily see that the grand coalition is the most efficient, the core in this game may not be fair to the players. Examples of unfair allocations that lie in the core abound both in the GT and SP literature [74], [75]. Moreover, drawing yet another analogy with the NE, there is no guarantee that a coalition game will have a core solution. Indeed, the core, as a set-valued solution, may be empty.

Nonetheless, the core is one of the most popular set-valued solution concepts in a coalition-form game which has led to many extensions. For instance, when dealing with a non-superadditive coalition-formation game with TU, we can redefine the core, based on the partition of $\mathcal{K}$ that maximizes the total utility, as follows:

$$\mathcal{O} = \left\{ x : \sum_{i \in \mathcal{K}} x_i = \max_{\pi \in \mathcal{P}} \sum_{\mathcal{C} \in \pi} v(\mathcal{C}) \text{ and } \sum_{i \in \mathcal{C}} x_i \geq v(\mathcal{C}) \ \forall \ \mathcal{C} \subseteq \mathcal{K} \right\}, \tag{43}$$

where $\mathcal{P}$ is the set of all possible partitions of $\mathcal{K}$ and $\pi$ is one such partition or coalition structure. Recall that the partition of the set $\mathcal{K}$ is a collection of disjoint subsets whose union would span the entire set $\mathcal{K}$. Thus, the partition constitutes the coalitions that are expected to form in the system. Essentially, the difference between (42) and (43) is that in (42) the first core condition assumes that the sum of the individual payoffs is equal to the value of the grand coalition, which is guaranteed to form due to superadditivity. In contrast, in (43), due to the non-superadditive nature of the game, the grand coalition is not guaranteed to form. Consequently, the first condition of the core must now ensure that the sum of the individual payoffs must be equal to the sum of the values of all coalitions in the partition $\pi$ that maximizes the total system value. Thus, this coalition-formation core notion implies that, instead of investigating a stable grand coalition, one would seek an allocation that will stabilize the partition $\pi$ that maximizes the total social welfare of the system. This is particularly useful when coalition formation entails a cost, and, thus, the game is non-superadditive.

*2) The $\epsilon$-core:* One extension to the core is the $\epsilon$-core. This notion bears an analogy with the notion of approximate or $\epsilon$-equilibria in strategic-form games [64]. The basic idea is that the stability is not achieved exactly, but rather within an $\epsilon$-approximation neighborhood as follows:

$$\mathcal{S}_\epsilon = \left\{ x : \sum_{i \in \mathcal{K}} x_i = v(\mathcal{K}) \text{ and } \sum_{i \in \mathcal{C}} x_i \geq v(\mathcal{C}) - \epsilon \ \forall \ \mathcal{C} \subseteq \mathcal{K}, \ \epsilon \geq 0 \right\}. \tag{44}$$

Interestingly enough, the value of $\epsilon$ can be viewed as a quantification of the "overhead" for deviating from the core. This overhead is incurred on the deviation of every possible coalition. This bears a very interesting analogy to SP – what is the overhead required by a group of devices to deviate from the stability concept and will they be willing to incur this overhead. The above concept is also known as the weak $\epsilon$-core, which is used to then define the so-called strong $\epsilon$-core, where $\epsilon$ is divided between the members of a coalition, i.e., $\epsilon$ is substituted by $|\mathcal{C}| \cdot \epsilon$. In this case, the overhead $\epsilon$ is implicitly assumed to be equally divided between coalition members. The advantage of the $\epsilon$-core is that it may be easier to establish its existence as well as to develop algorithms that can reach it. This simply mimics the advantages of any approximate solution concept in GT. In SP, there have been some recent works (e.g., [76]) that explored the $\epsilon$-core as a suitable concept for investigating problems related to beamforming where the overhead of deviating from a certain beamforming strategy might be high enough to reach an $\epsilon$-core and, thus, avoiding the need to reach the more stringent core definition.

*3) The Shapley value:* The core and its variants constitute set-valued stability notions. In contrast, we can solve a coalition-form game using single-valued fairness notions. Single-valued solution concepts mainly associate with every coalition game $(\mathcal{K}, v)$ a *unique* payoff vector known as the solution or value of the game (which is different from the value of a coalition). One example of such notion is the NBS that was previously discussed. In fact, most single-valued notions follow an axiomatic approach: a set of pre-set properties that are imposed on the sought after payoff allocation in order to find a desirable solution. One popular such solution is the *Shapley value* [23]. For a TU coalition-formation game, the Shapley value assigns to every player the payoff $x_i$ given by

$$x_i = \sum_{\mathcal{C} \subseteq \mathcal{K} \setminus \{i\}} \frac{|\mathcal{C}|!(|\mathcal{K}| - |\mathcal{C}| - 1)!}{|\mathcal{K}|!} [v(\mathcal{C} \cup \{i\}) - v(\mathcal{C})]. \tag{45}$$

This allocation is interpreted as follows. In the event where the players join the grand coalition in an arbitrary order, the payoff allotted by the Shapley value to a player $i \in \mathcal{K}$ is the expected marginal contribution of player $i$ when it joins the grand coalition. In other words, the contribution of a player is given by an expected value, assuming a random order of joining of the players to the grand coalition which, in a superadditive game, is known to be the most efficient solution. Shapley showed that this solution is unique and it satisfies the following four axioms:

a) *efficiency axiom*: $\sum_{i \in \mathcal{K}} x_i = v(\mathcal{K})$.

b) *symmetry axiom*: if player $i$ and player $j$ are such that $v(\mathcal{C} \cup \{i\}) = v(\mathcal{C} \cup \{j\})$ for every coalition $\mathcal{C}$ not containing player $i$ and player $j$, then $x_i = x_j$.

c) *dummy axiom*: if player $i$ is such that $v(\mathcal{C}) = v(\mathcal{C} \cup \{i\})$ for every coalition $\mathcal{C}$ not containing $i$, then $x_i = 0$.

d) *additivity axiom*: If $u$ and $v$ are characteristic functions, then the Shapley value allotted to the game over $(u + v)$ is the sum of the values allotted to $u$ and $v$, separately.

The Shapley value provides some form of fairness to allocate the payoffs of a grand coalition. Similarly to the core, the Shapley value has led to many extended notions such as the envy-free fairness [75], the Banzhaf index [23], and the Harsanyi index [23]. All of these notions follow the steps of the Shapley value in that they utilize certain axioms and attempt to find a coalition-form solution that satisfies these axioms. However, none of these solutions is guaranteed to be stable. For example, often, the Shapley value will not lie in the core, if that core exists. Therefore, one important challenge for coalition-form games is to balance fairness and stability by combining notions of core and Shapley value.

**Remark 5.** *In summary, for solving coalition-form games, a myriad of solution concepts exists. These are split into two categories: single-valued and set-valued. The focus is mainly on stability and fairness. The exact notion of stability or fairness depends largely on the type of the game and the scenario being considered.*

Next, we will discuss some principle results from coalition-form games and, then, we will delve more into algorithmic implementation and practical applications in the SP domain.

### D. Main theorems

Unlike strategic-form games in which existence, efficiency, and uniqueness theorems are abundant, for coalition-form games, such results are sparse and often model-dependent. However, when dealing with the core, we can discuss two seminal results that relate to the existence of the core and its fairness.

The first main result in this regard is given through the Bondareva-Shapley theorem [23]. This theorem is concerned with coalition-form games that are *balanced*:

**Definition 15.** *(Balanced game) A coalition TU game is said to be* balanced *if and only if we have*

$$\sum_{\mathcal{C} \subseteq \mathcal{K}} \mu(\mathcal{C}) v(\mathcal{C}) \leq v(\mathcal{K}), \tag{46}$$

*for all non-negative weight collections* $\mu = (\mu(\mathcal{C}))_{\mathcal{C} \subseteq \mathcal{K}}$.

Here, $\mu$ is simply a group of weights in $[0, 1]$ that are assigned to each coalition $\mathcal{C} \subseteq \mathcal{K}$ such that $\sum_{\mathcal{C} \ni i} \mu(\mathcal{C}) = 1$, $\forall i \in \mathcal{K}$. The main idea behind a balanced game can be explained as follows. Assuming that every player $i$ has a unit of time that can be divided between all possible coalitions that $i$ can form. Every coalition $\mathcal{C}$ is active for a time period $\mu(\mathcal{C})$ if all players in $\mathcal{C}$ are active during that time. The payoff of this active coalition would then be $\mu(\mathcal{C}) v(\mathcal{C})$. Here, $\sum_{\mathcal{C} \ni i} \mu(\mathcal{C}) = 1$, $\forall i \in \mathcal{K}$, would then be a feasibility constraint on the players' time allocation. Consequently, a coalition-form game is balanced if there is no feasible allocation of time which can yield an overall utility that exceeds the value $v(\mathcal{K})$ of the grand coalition. Thus, for a TU balanced game, the following result holds.

**Theorem 6** ([73]). *(Bondareva-Shapley) The core of a game is non-empty if and only if the game is balanced.*

Although the Bondareva-Shapley theorem is a popular result for showing the non-existence of the core, its applicability in SP may be very limited, as the required balancedness is quite restrictive on the coalition value. In this respect, yet another interesting result is given for *convex* coalition-form games. A coalition game with TU is said to be convex if its value function satisfies:

$$v(\mathcal{C}_1) + v(\mathcal{C}_2) \leq v(\mathcal{C}_1 \cup \mathcal{C}_2) + v(\mathcal{C}_1 \cap \mathcal{C}_2) \ \forall \mathcal{C}_1, \mathcal{C}_2 \subseteq \mathcal{K} \tag{47}$$

By observing (47), we can view directly its similarity with supermodular games, introduced in Sec. II-C. Now, supermodularity is defined with respect to subsets, rather than vectors in the Euclidean space. We note that the convexity conditions can also be written as follows:

$$v(\mathcal{C}_1 \cup \{i\}) - v(\mathcal{C}_1) \leq v(\mathcal{C}_2 \cup \{i\}) - v(\mathcal{C}_2) \tag{48}$$

whenever $\mathcal{C}_1 \subseteq \mathcal{C}_2 \subseteq \mathcal{K} \setminus \{i\}$. This can be explained as follows. A game is convex if and only if the marginal contribution of each player to a coalition is non-decreasing with respect to set inclusion. For a convex game, we can state the following theorem:

**Theorem 7** ([73]). *For a convex coalition-form game, the core is non-empty and the Shapley value lies in the core.*

This theorem provides a strong result that combines both stability and fairness. Indeed, for a convex game, the Shapley value is in the core and thus provides both stability and fairness. Although we stated the theorem here for TU games, it can also be extended to NTU games.

## V. Algorithms for coalition-form games

One key design challenge in coalition-form games is that of developing algorithms for characterizing and finding a suitable stable or fair solution. This is in general analogous with the algorithmic aspects of non-cooperative games where learning is needed to reach a certain NE (Sec. III). In this respect, here, we discuss two algorithmic aspects: *i)* finding a stable or a fair distribution for canonical games, and *ii)* characterizing stable partitions for coalition-formation games.

### A. Canonical games

For canonical games, the most important solution concept is the core and its variants. Despite being a strongly stable solution concept, computing the core can be relatively complex. In particular, in order to compute the core directly from the definition, we must solve the following linear program:

$$\text{minimize}_{\boldsymbol{x}} \quad \sum_{i \in \mathcal{K}} x_i \quad \text{s.t.} \quad \sum_{i \in \mathcal{C}} x_i \geq v(\mathcal{C}), \; \forall \mathcal{C} \subseteq \mathcal{K}. \tag{49}$$

Solving (49) allows us to find all the solutions that lie in the core, as ensured by the constraint. Clearly, solving the linear program in (49) will require handling $2^{\mathcal{K}}$ constraints, which will grow exponentially as the number of players increases. While no generic rule exists for overcoming this complexity, we can exploit some properties of the game or application being sought. On one hand, we can use theorems such as the Bondareva-Shapley theorem or the convexity of the game to establishing the existence and non-emptiness of the core (Sec. IV-D). On the other hand, for a given coalition-form game structure, we can evaluate the membership of known payoff division rules, such as the bargaining solution or a proportional fair division, in the core. Here, checking whether a certain allocation belongs in the core essentially becomes simpler than deriving all the solutions that are in the core.

Regarding the Shapley value, we can also observe a similar complexity limitation: computing the Shapley value via (45) calls for going again through all the possible coalitions. However, we note that, recently, some approximations for the Shapley value have been developed that allows us to compute it with reduced complexity. A popular approach in this context relies on the use of the *multi-linear extension* method proposed by Owen [77] for a special class of games known as voting games. The basic idea is to observe that in (45) the term inside the summation is the Beta function, which can then be used to convert the Shapley value computation into a probability computation which is then approximated by exploiting some properties of voting games. Other approaches to approximate or to reduce the computational time of the Shapley value are surveyed in [78].

### B. Coalition-formation games

Deriving suitable solutions for coalition-formation games is more challenging than the canonical case as it requires to jointly compute the payoff and the coalitional structure or network partition that will form. For example, computing coalitional structure that lie in the core, as per the definition in (43), can be highly complex, as it requires to look over all partitions of a set – which grow exponentially. However, some approaches using Markov chains or other related ideas have been proposed in [79], [80], which were proven to work well for reasonably large games.

However, in practical SP applications, we must trade off the strength of the core stability for the complexity of finding this solution. One baseline approach for a generic coalition-formation algorithm would consist of two key steps: *i)* define a rule using which a player may decide to join or leave a coalition and *ii)* for the TU case, adopt a proper payoff allocation rule (e.g., the Shapley value, proportional fair, etc.) that is to be applied at the level of any formed coalition.

Regarding the coalition-formation rule, a number of approaches have been proposed within the SP community (e.g., [12], [74], [76], [81]). Among them, the most popular ones are the merge and split rules, defined as follows ($\triangleright$ is a preference relation, discussed below):

- **Merge rule:** A group of coalitions $\{\mathcal{C}_1, \ldots, \mathcal{C}_p\}$ would merge into a single coalition $\cup_{k=1}^{p}\mathcal{C}_k$ if and only if $\cup_{k=1}^{p}\mathcal{C}_k \triangleright \{\mathcal{C}_1, \ldots, \mathcal{C}_p\}$,

- **Split rule:** A coalition $\cup_{k=1}^{p}\mathcal{C}_k$ will split into a smaller group of coalitions $\{\mathcal{C}_1, \ldots, \mathcal{C}_p\}$ if and only if $\{\mathcal{C}_1, \ldots, \mathcal{C}_p\} \triangleright \cup_{k=1}^{p}\mathcal{C}_k$.

Here, the preference relation $\triangleright$ can be defined based on the application being studied. A popular preference relations is the so-called *Pareto order*, whereby the merge or split rule would apply if at least one player improves its payoff via merge or split, without hurting the payoff of any other player. In other words, given the current payoff vector $y$ of all players involved in a merge or split rule, the merge or split occurs when the vector $x$ of the payoffs of all involved players is such that $x \geq y$ with at least one element $x_i$ of $x$ such that $x_i > y_i$. Essentially, this is reminiscent of the Pareto dominance rule used in non-cooperative games (Sec. II-B).

The advantages of using merge-and-split based algorithms include: *i)* guaranteed convergence to a stable, merge-and-split proof coalition structure after a finite number of iterations, *ii)* convergence is ensured irrespective of the starting point of the network, and *iii)* the order of merge or split will not impact convergence. Another major advantage of using merge-and-split based algorithms includes the fact that, irrespective of implementation, such algorithms will reach the so-called $\mathbb{D}_c$-stable partition, when such a partition exists. The $\mathbb{D}_c$-stable partition is a partition that: *i)* is strongly stable in the sense that no group of coalitions can do better by breaking away from this partition and *ii)* when using the Pareto order as a preference relation, is PO. Therefore, merge-and-split can reach such an optimal and strongly stable partition if it exists.[8]

### C. A case study: Coalition formation for collaborative target detection

One SP application in which the coalition-form can be applied is that of collaborative target detection. For example, in radar systems, a number of monitoring stations (MSs) can collaborate to detect a certain target of interest at a given location. Such stations can be located at different points in the network and, thus, their view on the target will be different. Here, we assume that the target is a wireless device that is transmitting a certain signal which must be detected. One major challenge in this scenario is the hidden terminal problem – due to fading and path loss some MSs may receive a weaker signal from the target, thus hindering their detection performance.

To avoid this problem, collaborative target detection (CTD) can be used. The basic idea is that MSs can share their individual detection results and, then, make a collective decision on the absence or presence of a target at a given location. By collaborating, the MSs can exploit the diversity of their observations to improve detection decisions. However, although CTD can improve the probability of detecting the target as the number of

---

[8]The existence of a $\mathbb{D}_c$-stable partition is highly application-dependent and the condition for existence will depend on the domain being studied.

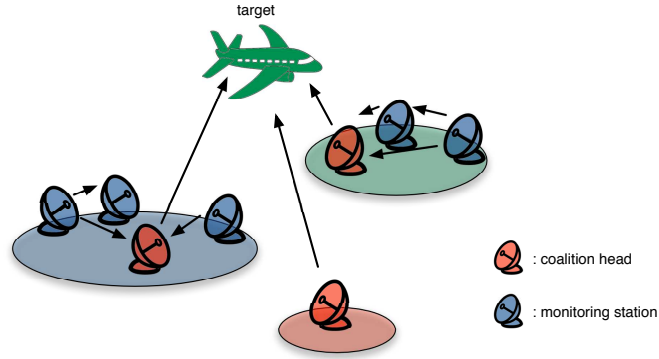

target

: coalition head

: monitoring station

Fig. 13: Distributed collaborative target detection as a coalition game.

collaborating MSs increases, collaboration can lead to an increasing probability of false alarm – the probability that a target is detected while it is not there. The tradeoff between probability of detection and probability of false alarm, as a function of the number of collaborating MSs, motivates the development of a coalition-form games in which the MSs can dynamically decide on how to collaborate while improving probability of detection and maintaining a tolerable false alarm level.

As shown in Fig. 13, we consider a coalition game between a set $\mathcal{K}$ of MSs that are seeking to cooperate in order to improve CTD performance. Since cooperation here entails a cost (in terms of increased false alarm), the game, in general, cannot be superadditive and, thus, it is classified as a coalition-formation game. In this game, each coalition $\mathcal{C}$ of secondary users (SUs) will be optimizing the following value function:

$$v(\mathcal{C}) = Q_{d,\mathcal{C}} - C(Q_{f,\mathcal{C}}, \alpha_{\mathcal{C}}) \tag{50}$$

where $Q_{d,\mathcal{C}}$ is the collaborative probability of detection and $C(\cdot)$ is a cost function of the collaborative false alarm level $Q_{f,\mathcal{C}}$ and the target false alarm constraint $\alpha_{\mathcal{C}}$. In this model, each coalition $\mathcal{C}$ will have a coalition head that will collect the detection results and fuse them in order to make a collective coalition decision.[9] Here, we notice that (50) is a probabilistic utility and, thus, it cannot be transferred between the members of $\mathcal{C}$. As a result, the CTD coalition-formation game is an NTU game with a special property: the payoff $x_i$ of each member $i$ of $\mathcal{C}$ is simply equal to $v(\mathcal{C})$, since this value is a collective result, i.e., we assume that all players in a coalition abide by the entire coalition decision.

Given the utility and involved tradeoffs, a merge-and-split algorithm based on the Pareto order can be proposed, as shown in [81] to find and characterize stable partitions. In Fig. 14, we show a snapshot of the network structure resulting from a merge-and-split collaborative spectrum sensing (CSS) algorithm (dashed line) as well as from a centralized approach (solid line) for seven randomly deployed MSs. We notice that the partitions resulting from both approaches are comparable, with neighboring MSs cooperating for improving spectrum sensing. However, this example allows us to highlight the difference between a distributed, coalition-formation game approach, in which each MSs makes its own CTD decision, and a centralized optimization approach, in which the MSs have no say in the coalition-formation process. In particular, from Fig. 14, we can see that for the game solution, MS4 is part of coalition $\{1, 2, 4, 6\}$, while for the centralized approach MS4 is member of $\{3, 4, 5\}$. This difference stems from the fact that, in the distributed case, MS4 acts selfishly while aiming at improving its own utility. In fact, by merging with $\{3, 5\}$, MS4 achieves a utility of $0.9859$ with a probability of detection of $0.9976$ whereas by merging with $\{1, 2, 6\}$ its utility will be $0.9957$ with a probability of detection of $0.99901$. Thus, in a coalition-based solution, MS4 prefers to merge with $\{1, 2, 6\}$ rather than with $\{3, 5\}$ regardless of the socially optimal partition.

[9]The fusion rule used will impact the way in which $Q_{d,\mathcal{C}}$ and $Q_{f,\mathcal{C}}$ are computed. However, it will not affect the way the game is formulated.

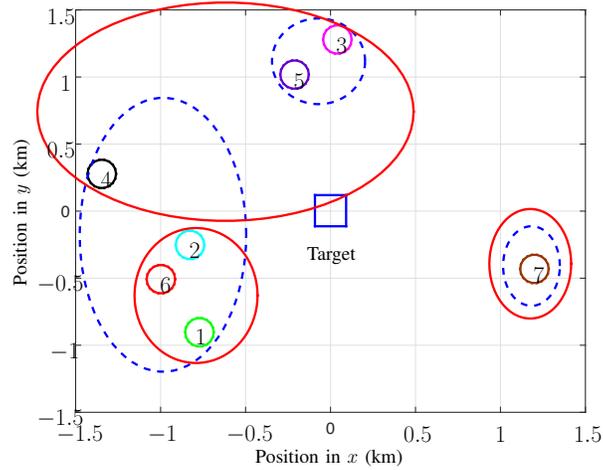

Fig. 14: Final coalition structure from both distributed (dashed line) and centralized (solid line) collaborative target detection for $K = 7$ MSs.

In summary, the use of a coalition-formation game for CTD can also yield significant gains in terms of the probability of detection, while maintaining a required false alarm level and without the need for a centralized optimization solution. Building on these results, we can develop a broad range of applications that adopt the coalition-form games for SP problems. For example, the aforementioned model for CSS is extended in [82] to the case in which an MS can belong simultaneously to multiple coalitions. In this regard, [82] shows that the merge-and-split algorithm can be extended to handle the cases of OCF games.

## VI. CONCLUSIONS

In this tutorial, we have provided a holistic view on the use of game-theoretic techniques in signal processing for networks. Particular emphasis has been given to games in strategic and coalitional forms. The key components of such games have been introduced and discussed while providing a signal-processing-oriented view on the various types of games. Some of the primary differences and properties of strategy-form and coalition-form games are summarized in Table III. Then, we have developed the main solution concepts and discussed the various advantages and drawbacks within signal processing domains. More importantly, this tutorial has attempted to provide an in-depth discussion on the connections between game theory and algorithmic aspects of signal processing techniques. The applications discussed range from traditional communication problems to modern-day signal processing problems such as cognitive radio and wireless sensor networks. Overall, this tutorial is expected to provide a comprehensive, self-contained reference on the challenges and opportunities for adopting game theory in signal processing, as well as to locate specific references either in applications or theory.

Table III: Strategic form vs Coalition form.

| | **Strategic form** | **Coalition form** |
|---|---|---|
| *Components* | players, actions, per-player utility | players, coalition value, per-player utility |
| *Primary player strategy* | choose a parameter to optimize | choose a coalition membership |
| *Primary player objective* | optimize individual utility | optimize individual utility (while part of a coalition) |
| *Game objectives* | find an equilibrium | find stable coalitions and fair allocations |
| *Main types* | static, dynamic | TU, NTU, canonical, coalition formation, CF, PF |
| *Communication* | no communication between players | players can form agreements and communicate |
| *Main solution concept* | NE – no player can unilaterally deviate | stable partition – no coalition can deviate |
| *Baseline algorithms* | learning based | merge-and-split based |
| *Primary application* | distributed optimization | optimized cooperation, resource distribution |

## REFERENCES

[1] S. Kassam and H. Poor, "Robust signal processing for communication systems," *IEEE Commun. Mag.*, vol. 21, no. 1, pp. 20–28, Jan. 1983.

[2] A. Cohen and A. Lapidoth, "The Gaussian watermarking game," *IEEE Trans. Information Theory*, vol. 48, no. 6, pp. 1639–1667, Jun. 2002.

[3] C. Saraydar, N. B. Mandayam, and D. Goodman, "Efficient power control via pricing in wireless data networks," *IEEE Trans. Commun.*, vol. 50, no. 2, pp. 291–303, Feb. 2002.

[4] E. Larsson and E. Jorswieck, "Competition versus cooperation on the MISO interference channel," *IEEE J. Sel. Areas Commun.*, vol. 26, no. 7, pp. 1059–1069, Sep. 2008.

[5] G. Scutari, D. Palomar, and S. Barbarossa, "The MIMO iterative waterfilling algorithm," *IEEE Trans. Signal Process.*, vol. 57, no. 5, pp. 1917–1935, May 2009.

[6] Y. Wu and K. Liu, "An information secrecy game in cognitive radio networks," *IEEE Trans. Information Forensics and Security*, vol. 6, no. 3, pp. 831–842, Sep. 2011.

[7] B. Wang, K. Liu, and T. Clancy, "Evolutionary cooperative spectrum sensing game: how to collaborate?" *IEEE Trans. Commun.*, vol. 58, no. 3, pp. 890–900, Mar. 2010.

[8] F. Wang, M. Krunz, and S. Cui, "Price-based spectrum management in cognitive radio networks," *IEEE J. Sel. Topics Signal Process.*, vol. 2, no. 1, pp. 74–87, Feb. 2008.

[9] H. Park and M. van der Schaar, "Bargaining strategies for networked multimedia resource management," *IEEE Trans. Signal Process.*, vol. 55, no. 7, pp. 3496–3511, Jul. 2007.

[10] A. Chakraborty and J. Duncan, "Game-theoretic integration for image segmentation," *IEEE Trans. Pattern Analysis Machine Intelligence*, vol. 21, no. 1, pp. 12–30, Jan. 1999.

[11] B. Ibragimov, B. Likar, F. Pernus, and T. Vrtovec, "A game-theoretic framework for landmark-based image segmentation," *IEEE Trans. Medical Imaging*, vol. 31, no. 9, pp. 1761–1776, Sep. 2012.

[12] H. He, A. Subramanian, X. Shen, and P. Varshney, "A coalitional game for distributed estimation in wireless sensor networks," in *Proc. IEEE Int. Conf. Acoustics, Speech and Signal Processing (ICASSP)*, Vancouver, Canada, May 2013.

[13] C. Jiang, Y. Chen, and K. Liu, "Distributed adaptive networks: A graphical evolutionary game-theoretic view," *IEEE Trans. Signal Process.*, vol. 61, no. 22, pp. 5675–5688, Nov. 2013.

[14] K. Han and A. Nehorai, "Joint frequency-hopping waveform design for MIMO radar estimation using game theory," in *Proc. IEEE Radar Conf.*, Ottawa, Canada, Apr. 2013.

[15] A. Mukherjee and A. Swindlehurst, "Jamming games in the MIMO wiretap channel with an active eavesdropper," *IEEE Trans. Signal Process.*, vol. 61, no. 1, pp. 82–91, Jan. 2013.

[16] X. Song, P. Willett, S. Zhou, and P. Luh, "The MIMO radar and jammer games," *IEEE Trans. Signal Process.*, vol. 60, no. 2, pp. 687–699, Feb. 2012.

[17] A. Moragrega, P. Closas, and C. Ibars, "Supermodular game for power control in TOA-based positioning," *IEEE Trans. Signal Process.*, vol. 61, no. 12, pp. 3246–3259, Dec. 2013.

[18] K. S. Narendra and M. A. L. Thathachar, *Learning Automata: An Introduction.* Harlow, UK: Prentice Hall, 1989.

[19] G. Scutari, D. Palomar, F. Facchinei, and J.-S. Pang, "Convex optimization, game theory, and variational inequality theory," *IEEE Signal Process. Mag.*, vol. 27, no. 3, pp. 35–49, May 2010.

[20] J. C. Harsanyi and R. Selten, *A General Theory of Equilibrium Selection in Games.* Cambridge, MA: MIT Press, Mar. 2003.

[21] J. Y. Halpern, "A computer scientist looks at game theory," *Games and Economic Behavior*, vol. 45, no. 1, pp. 114–131, Oct. 2003.

[22] S. Lasaulce and H. Tembine, *Game Theory and Learning for Wireless Networks: Fundamentals and Applications.* Waltham, MA: Academic Press, 2011.

[23] Z. Han, D. Niyato, W. Saad, T. Başar, and A. Hjørungnes, *Game Theory in Wireless and Communication Networks: Theory, Models and Applications.* Cambridge, UK: Cambridge Univ. Press, 2011.

[24] E. Jorswieck, E. Larsson, M. Luise, and H. Poor, "Game theory in signal processing and communications [Guest Editorial]," *IEEE Signal Process. Mag*, vol. 26, no. 5, pp. 17, 132, Sep. 2009.

[25] E. Altman, T. Boulogne, R. El-Azouzi, T. Jiménez, and L. Wynter, "A survey on networking games in telecommunications," *Comput. Oper. Res.*, vol. 33, no. 2, pp. 286–311, Feb. 2006.

[26] N. Mandayam, S. Wicker, J. Walrand, T. Başar, J. Huang, and D. Palomar, "Game theory in communication systems [Guest Editorial]," *IEEE J. Sel. Areas Commun.*, vol. 26, no. 7, pp. 1042–1046, Sep. 2008.

[27] G. Bacci, L. Sanguinetti, and M. Luise, "Understanding game theory via wireless power control," *IEEE Signal Process. Mag*, vol. 32, no. 4, pp. 132–137, Jul. 2015.

[28] E. Björnson, E. Jorswieck, M. Debbah, and B. Ottersten, "Multi-objective signal processing optimization: The way to balance conflicting metrics in 5G systems," *IEEE Signal Process. Mag.*, vol. 31, no. 6, pp. 14–23, Nov. 2014.

[29] J. F. Nash, "Non-cooperative games," *Annals of Mathematics*, vol. 54, no. 2, pp. 286–295, 1951.

[30] R. J. Aumann, "Acceptable points in general cooperative *n*-person games," in *Contributions to the Theory of Games*, ser. Annals of Mathematics Studies 40, R. D. Luce and A. W. Tucker, Eds. Princeton, NJ: Princeton Univ. Press, 1959, vol. IV, pp. 287–324.

[31] D. Fudenberg and J. Tirole, *Game Theory.* Cambridge, MA: MIT Press, 1991.

[32] M. J. Osborne and A. Rubinstein, *A Course in Game Theory.* Cambridge, MA: MIT Press, 1994.

[33] R. Mazumdar, L. Mason, and C. Douligeris, "Fairness in network optimal flow control: Optimality of product forms," *IEEE Trans. Commun.*, vol. 39, no. 5, pp. 775–782, May 1991.

[34] C. Touati, E. Altman, and J. Galtier, "Generalized Nash bargaining solution for bandwidth allocation," *Computer Networks*, vol. 50, no. 17, pp. 3242–3263, Dec. 2006.

[35] C. H. Papadimitriou, "Algorithms, games, and the Internet," in *Proc. ACM Annual Symp. Theory of Computing*, Heraklion, Greece, Jul. 2001.

[36] J. R. Correa, A. S. Schulz, and N. E. Stier-Moses, "Selfish routing in capacitated networks," *Math. Operations Res.*, vol. 29, no. 4, pp. 961–976, Nov. 2004.

[37] R. B. Myerson, "Mechanism design," in *The New Palgrave Dictionary of Economics*, 2nd ed., S. N. Durlauf and L. E. Blume, Eds. London, U.K.: Palgrave Macmillan, 2008.

[38] J. F. Nash, "The bargaining problem," *Econometrica*, vol. 18, no. 2, pp. 155–162, Apr. 1950.

[39] H. Boche and M. Schubert, "Nash bargaining and proportional fairness for wireless systems," *IEEE/ACM Trans. Networking*, vol. 17, no. 5, pp. 1453–1466, Oct. 2009.

[40] Z. Liu, Z. Wu, P. Liu, H. Liu, and Y. Wang, "Layer bargaining: multicast layered video over wireless networks," *IEEE J. Sel. Areas Commun.*, vol. 28, no. 3, pp. 445–455, Apr. 2010.

[41] O. Compte and P. Jehiel, "The coalitional nash bargaining solution," *Econometrica*, vol. 78, no. 5, pp. 1593–1623, Sep. 2010.

[42] E. Larsson and E. Jorswieck, "Competition versus cooperation on the MISO interference channel," *IEEE J. Sel. Areas Commun.*, vol. 26, no. 7, pp. 1059–1069, Sep. 2008.

[43] S. Gogineni and A. Nehorai, "Game theoretic design for polarimetric MIMO radar target detection," *Signal Processing*, vol. 92, no. 5, pp. 1281 – 1289, May 2012.

[44] T. M. Cover and J. A. Thomas, *Elements of Information Theory*, 2nd ed.   New York: J. Wiley & Sons, 2006.

[45] E. V. Belmega, S. Lasaulce, and M. Debbah, "Power allocation games for MIMO multiple access channels with coordination," *IEEE Trans. Wireless Commun.*, vol. 8, no. 6, pp. 3182–3192, Jun. 2009.

[46] D. M. Topkis, *Supermodularity and Complementarity*.   Princeton, NJ: Princeton Univ. Press, 1998.

[47] S. Boyd and L. Vandenberghe, *Convex Optimization*.   Cambridge, UK: Cambridge Univ. Press, 2002.

[48] P. Mertikopoulos, E. V. Belmega, A. Moustakas, and S. Lasaulce, "Distributed learning policies for power allocation in multiple access channels," *IEEE J. Sel. Areas Commun.*, vol. 30, no. 1, pp. 96–106, Jan. 2012.

[49] M. Piezzo, A. Aubry, S. Buzzi, A. D. Maio, and A. Farina, "Non-cooperative code design in radar networks: A game-theoretic approach," *EURASIP J. Advances Signal Processing*, vol. 2013:63, 2013.

[50] Y. Yang, F. D. Rubio, G. Scutari, and D. P. Palomar, "Multi-portfolio optimization: A potential game approach," *IEEE Trans. Signal Process.*, vol. 61, no. 22, pp. 5590–5602, Nov. 2013.

[51] J. Marden, G. Arslan, and J. Shamma, "Cooperative control and potential games," *IEEE Trans. Systems, Man, and Cybernetics, Part B: Cybernetics*, vol. 39, no. 6, pp. 1393–1407, Dec. 2009.

[52] J. C. Harsanyi, "Games with incomplete information played by Bayesian players," *Management Science*, vol. 14, pp. 159–182, 320–334, 486–502, 1967-1968.

[53] G. Bacci and M. Luise, "A game-theoretic perspective on code synchronization for CDMA wireless systems," *IEEE J. Sel. Areas Commun.*, vol. 30, no. 1, pp. 107–118, Jan. 2012.

[54] V. Krishnamurthy and H. Poor, "Social learning and Bayesian games in multiagent signal processing: How do local and global decision makers interact?" *IEEE Signal Process. Mag.*, vol. 30, no. 3, pp. 43–57, May 2013.

[55] D. G. Harper, "Competitive foraging in mallards: 'Ideal free' ducks," *Anim. Behav.*, vol. 30, no. 2, pp. 575–584, May 1982.

[56] G. H. Golub and C. F. Van Loan, *Matrix Computations*, 3rd ed.   Baltimore, MD: Johns Hopkins Univ. Press, 1996.

[57] S. Lloyd, "Least squares quantization in PCM," *IEEE Trans. Inf. Theory*, vol. 28, no. 2, pp. 129–137, Mar. 1982.

[58] B. Larrousse, O. Beaude, and S. Lasaulce, "Crawford-sobel meet Lloyd-Max on the grid," in *IEEE International Conference on Acoustics, Speech and Signal Processing (ICASSP)*, Florence, Italy, May 2014, pp. 6127–6131.

[59] D. Niyato and E. Hossain, "Competitive spectrum sharing in cognitive radio networks: A dynamic game approach," *IEEE Trans. Wireless Commun.*, vol. 7, no. 7, pp. 2651–2660, Jul. 2008.

[60] R. D. Yates, "A framework for uplink power control in cellular radio systems," *IEEE J. Select. Areas Commun.*, vol. 13, no. 9, pp. 1341–1347, Sep. 1995.

[61] L. Gan, U. Topcu, and S. H. Low, "Optimal decentralized protocol for electric vehicle charging," *IEEE Trans. Power Systems*, vol. 28, no. 2, pp. 940–951, May 2013.

[62] G. Scutari, D. Palomar, and S. Barbarossa, "Optimal linear precoding strategies for wideband noncooperative systems based on game theory part I: Nash equilibria," *IEEE Trans. Signal Process.*, vol. 56, no. 3, pp. 1230–1249, Mar. 2008.

[63] W. Yu, G. Ginis, and J. Cioffi, "Distributed multiuser power control for digital subscriber lines," *IEEE J. Sel. Areas Commun.*, vol. 20, no. 5, pp. 1105–1115, Jun. 2002.

[64] L. Rose, S. Lasaulce, S. Perlaza, and M. Debbah, "Learning equilibria with partial information in decentralized wireless networks," *IEEE Commun. Mag.*, vol. 49, no. 8, pp. 136–142, Aug. 2011.

[65] G. W. Brown, "Iterative solutions of games by fictitious play," in *Activity Analysis of Production and Allocation*, T. C. Koopmans, Ed.   New York: Wiley, 1951, pp. 374–376.

[66] R. R. Bush and F. Mosteller, *Stochastic Models of Learning*, 1st ed.   New York: John Wiley & Sons, 1955.

[67] S. Hart and A. Mas-Colell, "A simple adaptive procedure leading to correlated equilibrium," *Econometrica*, vol. 68, no. 5, pp. 1127–1150, Sep. 2000.

[68] M. Bennis, S. M. Perlaza, and M. Debbah, "Learning coarse correlated equilibrium in two-tier wireless networks," in *Proc. IEEE Int. Conf. Communications (ICC)*, Ottawa, Canada, Jun. 2012.

[69] L. Rose, S. Perlaza, and M. Debbah, "On the Nash equilibria in decentralized parallel interference channels," in *Proc. IEEE Int. Conf. Commun. (ICC)*, Kyoto, Japan, Jun. 2011.

[70] R. Olfati-Saber, J. Fax, and R. Murray, "Consensus and cooperation in networked multi-agent systems," *Proc. IEEE*, vol. 95, no. 1, pp. 215–233, Jan. 2007.

[71] E. Semsar-Kazerooni and K. Khorasani, "Multi-agent team cooperation: A game theory approach," *Automatica*, vol. 45, no. 10, pp. 2205–2213, Oct. 2009.

[72] R. J. Aumann and B. Peleg, "Von Neumann-Morgenstern solutions to cooperative games without side payments," *Bull. Amer. Math. Soc.*, vol. 66, no. 3, pp. 173–179, 1960.

[73] G. Owen, *Game Theory*, 3rd ed.   London, UK: Academic Press, 1995.

[74] S. Mathur, L. Sankaranarayanan, and N. Mandayam, "Coalitions in cooperative wireless networks," *IEEE J. Sel. Areas Commun.*, vol. 26, no. 7, pp. 1104–1115, Sep. 2008.

[75] R. La and V. Anantharam, "A game-theoretic look at the Gaussian multiaccess channel," in *Proc. DIMACS Workshop on Network Information Theory*, Piscataway, NJ, Mar. 2003.

[76] R. Mochaourab and E. Jorswieck, "Coalitional games in MISO interference channels: Epsilon-core and coalition structure stable set," *IEEE Trans. Signal Process.*, vol. 62, no. 24, pp. 6507–6520, Dec. 2014.

[77] G. Owen, "Multilinear extensions of games," *Management Science*, vol. 18, no. 5, pp. 64–79, Jan. 1972.

[78] S. S. Fatima, M. Woodridge, and N. R. Jennings, "A linear approximation method for the Shapley value," *Artificial Intelligence*, vol. 172, no. 14, pp. 1673–1699, Sep. 2008.

[79] T. Arnold and U. Schwalbe, "Dynamic coalition formation and the core," *J. Economic Behavior and Organization*, vol. 49, no. 3, pp. 363–380, Nov. 2002.

[80] D. Niyato, P. Wang, W. Saad, and A. Hjørungnes, "Controlled coalitional games for cooperative mobile social networks," *IEEE Trans. Veh. Technol.*, vol. 60, no. 4, pp. 1812–1824, May 2011.

[81] W. Saad, Z. Han, T. Başar, M. Debbah, and A. Hjørungnes, "Coalition formation games for collaborative spectrum sensing," *IEEE Trans. Veh. Technol.*, vol. 60, no. 1, pp. 276–297, Jan. 2011.

[82] T. Wang, L. Song, Z. Han, and W. Saad, "Distributed cooperative sensing in cognitive radio networks: An overlapping coalition formation approach," *IEEE Trans. Commun.*, vol. 62, no. 9, pp. 3144–3160, Sep. 2014.